\documentclass[pra,reprint,aps,superscriptaddress,showpacs,showkeys,amsmath,amssymb]{revtex4-2}
\usepackage{graphicx,xcolor,dcolumn,bm,hyperref,times}
\usepackage[T1]{fontenc}
\definecolor{linkblue}{rgb}{0, 0, 1}
\hypersetup{
	colorlinks=true,
	linkcolor=linkblue,
	citecolor=linkblue,
	urlcolor=linkblue,
}

\begin{document}

\title{Theory of the monochromatic advanced-wave picture and applications in biphoton optics}

\author{Yi~Zheng}
\affiliation{CAS Key Laboratory of Quantum Information, University of Science and Technology of China, Hefei 230026, China}
\affiliation{Anhui Province Key Laboratory of Quantum Network, University of Science and Technology of China, Hefei 230026, China}
\affiliation{CAS Center for Excellence in Quantum Information and Quantum Physics, University of Science and Technology of China, Hefei 230026, China}

\author{Jin-Shi~Xu}
\email{jsxu@ustc.edu.cn}
\affiliation{CAS Key Laboratory of Quantum Information, University of Science and Technology of China, Hefei 230026, China}
\affiliation{Anhui Province Key Laboratory of Quantum Network, University of Science and Technology of China, Hefei 230026, China}
\affiliation{CAS Center for Excellence in Quantum Information and Quantum Physics, University of Science and Technology of China, Hefei 230026, China}
\affiliation{Hefei National Laboratory, University of Science and Technology of China, Hefei 230088, China}

\author{Chuan-Feng~Li}
\email{cfli@ustc.edu.cn}
\affiliation{CAS Key Laboratory of Quantum Information, University of Science and Technology of China, Hefei 230026, China}
\affiliation{Anhui Province Key Laboratory of Quantum Network, University of Science and Technology of China, Hefei 230026, China}
\affiliation{CAS Center for Excellence in Quantum Information and Quantum Physics, University of Science and Technology of China, Hefei 230026, China}
\affiliation{Hefei National Laboratory, University of Science and Technology of China, Hefei 230088, China}

\author{Guang-Can~Guo}
\affiliation{CAS Key Laboratory of Quantum Information, University of Science and Technology of China, Hefei 230026, China}
\affiliation{Anhui Province Key Laboratory of Quantum Network, University of Science and Technology of China, Hefei 230026, China}
\affiliation{CAS Center for Excellence in Quantum Information and Quantum Physics, University of Science and Technology of China, Hefei 230026, China}
\affiliation{Hefei National Laboratory, University of Science and Technology of China, Hefei 230088, China}

\date{\today}

\begin{abstract}
Klyshko's advanced-wave picture (AWP) is mainly interpreted by replacing the nonlinear crystal producing biphotons via spontaneous parametric down-conversion (SPDC) by a mirror in quantum imaging protocols with thin crystals, where the biphotons are perfectly correlated in position at the crystal. To better explain the biphoton spatial states produced by arbitrary crystals and pump beams, we develop a formal theory of AWP with monochromatic lights that the conditional wave function of one photon is calculated by propagation, multiplication, and another propagation. The case of more general photon postselection or no detection and the inclusion of polarization are studied. Then, we explain the form of the biphoton state from SPDC with a bulk crystal and its free-space propagation. By treating the biphoton wave function as an impulse response function of a classical optical setup, we analyze quantum imaging with undetected photons and quantum holography with polarization entanglement, where properties like the spatial resolution can be concisely deduced. This method can be employed to design nonlinear materials or novel quantum imaging techniques. Finally, we discuss Klyshko's original proposal beyond monochromatic lights with the Hong-Ou-Mandel effect as an example.
\end{abstract}

\maketitle

\section{Introduction}

The quantum optical field with photon entanglement \cite{Zhang:24} is distinct from its classical counterpart. The polarization, discrete spatial path, and time-bin degrees of freedom (DoF) of entangled photons and the squeezed state \cite{Walls1983} have been exploited in quantum foundation \cite{Shadbolt2014} and quantum information researches \cite{photonrev}. Here, we focus on the spatial DoF of the two-photon (biphoton) optical field, where spontaneous parametric down-conversion (SPDC) \cite{PhysRevA.31.2409,WALBORN201087,Schneeloch_2016,Karan_2020,PhysRevResearch.4.033098}, a quantum second-order nonlinear process, is mainly used for state creation. In addition, four-wave mixing with atomic ensembles can produce biphotons with much narrower spectral bands \cite{PhysRevLett.117.250501,sciadv.adf9161}. The form of the SPDC biphoton transverse spatial state (sometimes called the wave function \cite{Lundeen2011,Zheng2023,Zia2023}) and its propagation have been studied \cite{PhysRevA.53.2804,PhysRevA.57.3123,PhysRevA.62.043816,Abouraddy:02,PhysRevLett.92.127903,PhysRevLett.94.223601,PhysRevA.75.050101,PhysRevA.95.063836,Bhattacharjee2022,qshws}. Besides the continuous-variable Einstein-Podolsky-Rosen paradox \cite{EPRpaper,PhysRevLett.92.210403}, a prominent research field involving the spatial DoF is quantum imaging \cite{Moreau2019,lpor.201900097,Defienne2024}, which can be further categorized into counter-intuitive imaging, imaging against noise \cite{distill,sciadv.aay2652}, superresolution imaging \cite{PhysRevA.79.013827,Toninelli:19}, and enhanced interference \cite{PhysRevLett.85.2733,Ndagano2022}, phase \cite{Defienne2021,sciadv.abj2155,Black:23,Zia2023,science.adk7825}, or plenoptic imaging \cite{PhysRevLett.116.223602,app8101958,PhysRevApplied.21.024029}. There are two main counter-intuitive imaging protocols, ghost imaging (GI) \cite{ghostimaging,shihlzgx} and quantum imaging with undetected photons (QIUP) \cite{Lemos2014,PhysRevA.92.013832,RevModPhys.94.025007,BarretoLemos:22,qute.202300353}. In GI, the beam passing through the object is detected by a bucket detector, and the image is obtained by triggering a spatially resolved detector at the other path. GI with position-entangled photon pairs was originally proposed and realized, and it was later found that a thermal light field can also achieve GI due to its second-order correlation \cite{PhysRevLett.94.063601}. In QIUP, the beam passing through the object is discarded, while the object including its phase distribution can still be imaged because it determines the spatial coherence of two other beams.

When studying quantum imaging protocols using SPDC, the biphoton joint probability distribution (JPD) needs to be theoretically analyzed. However, the biphoton phenomena are difficult to calculate and understand. In 1988, Klyshko suggested the advanced-wave picture (AWP) \cite{Klyshko_1988,KLYSHKO1988133} which originally involves the time direction of the optical waves (see Sec.\ \ref{discusssec}). When applied to GI, its \emph{better-known meaning} is that the conditional spatial state of photon 2 when photon 1 is postselected to a certain position is the same as the amplitude of the reflected light from a point source at the postselection position in a classical setup where the nonlinear crystal (NLC) is replaced by a mirror. This assumes the pump beam is a normally incident plane wave and the NLC is sufficiently thin. When the wavefront of the pump beam is curved, so is the mirror \cite{Belinskii1994,PhysRevA.53.2804,WALBORN201087}. The AWP was experimentally verified by comparing the ghost image and the classical reflection result \cite{Klyshkoexp} and was generalized to stimulated parametric down-conversion \cite{PhysRevA.98.023850}. From the relation between the first- and second-order correlation of the thermal light and the van Cittert-Zernike theorem, GI with thermal light can also be unfolded, but the mirror is a phase conjugated one \cite{PhysRevA.62.043816,PhysRevA.71.013801}.

However, to our knowledge, related articles in quantum imaging mainly use the better-known meaning of the AWP as a tool, and a formal theory including its analysis of the impact of NLC thickness in quantum imaging is lacking. Also, the two photons are postselected to two points to use this theory. Although its extension to bucket detection of one photon in GI is easy, the photon may even be undetected in some protocols like QIUP. In this article, we start from the optical reciprocity, and present the general AWP theory for monochromatic lights including the polarization DoF and other postselection methods. Then, we analyze the form of SPDC biphoton wave function, describe the biphoton propagation in the free space, and illustrate the decisive factors of properties of QIUP and quantum holography with polarization entanglement, including the spatial resolutions and fields of view. Finally, we briefly discuss applications of the AWP including the numerical calculation of biphoton diffraction and the design of nonlinear crystals \cite{PhysRevLett.81.4136,Zhang:21,Rozenberg:22,adma.202313589}, present possible theoretical studies in the future including the generation of arbitrary biphoton spatial state, and mention the original proposal of AWP beyond monochromatic lights which can explain some quantum interference phenomena including the Hong-Ou-Mandel (HOM) effect \cite{PhysRevLett.59.2044}.

\section{Theory}

\subsection{Impulse response function of optical systems}\label{IRFoOS}

Before presenting the AWP theory, we need to clarify the concepts and notations used in this article. We use $\boldsymbol{\rho}=(x,y)$ and $\mathbf{r}=(x,y,z)=(\boldsymbol{\rho},z)$ to describe transverse and three-dimensional positions, respectively. In a linear, passive optical system which is insensitive to polarization and may be absorptive, the transformation of a monochromatic optical field can be described by the impulse response function $h(\boldsymbol{\rho},\boldsymbol{\rho}_0)$, where $\boldsymbol{\rho}_0,\boldsymbol{\rho}$ are points at the input and output plane respectively. For the input electric field complex amplitude $U_0(\boldsymbol{\rho})$, the output amplitude is
\begin{equation}\label{ampirf}
    U(\boldsymbol{\rho})=\int d\boldsymbol{\rho}_0U_0(\boldsymbol{\rho}_0)h(\boldsymbol{\rho},\boldsymbol{\rho}_0).
\end{equation}
If the optical system is a thin amplitude or phase object (or both) with the transmittance distribution $T(\boldsymbol{\rho})$ [$|T(\boldsymbol{\rho})|\leq1$], then $h(\boldsymbol{\rho},\boldsymbol{\rho}_0)=T(\boldsymbol{\rho}_0)\delta(\boldsymbol{\rho}-\boldsymbol{\rho}_0)$. If it is a linear, homogeneous, and isotropic (LHI) medium with the thickness $L$, refractive index $n$, and an infinite transverse size, we can let this shift-invariant system be $h(\boldsymbol{\rho},\boldsymbol{\rho}_0)=G_L(\boldsymbol{\rho}-\boldsymbol{\rho}_0)$, where $G_L(\boldsymbol{\rho})$ is the Green's function. From the Fresnel diffraction formula,
\begin{equation}\label{greeniso}
    G_L(\boldsymbol{\rho})=\frac{ne^{inkL}}{i\lambda L}\exp\left(i\frac{nk}{2L}|\boldsymbol{\rho}|^2\right),
\end{equation}
where $\lambda=2\pi c/\omega$ and $k=2\pi/\lambda=\omega/c$ are the wavelength and wave number of light with the angular frequency $\omega$ in the \emph{vacuum}. So, although the global phase $nkL$ is proportional to $n$, the equivalent diffraction distance in the vacuum is $L/n$. In a uniaxial birefringent crystal whose optical axis lies at the $x$-$z$ plane, the Green's function for the extraordinary (\emph{e}) light is approximately
\begin{equation}\label{greenbiref}
    G_{e,L}(\boldsymbol{\rho})\approx\frac{\eta e^{i\eta kL}}{i\lambda L}\exp\left(i\frac{\eta k}{2L}|\boldsymbol{\rho}+\alpha L\mathbf{e}_x|^2\right),
\end{equation}
where $\mathbf{e}_x$ is the unit vector toward the $x$ axis. See Appendix \ref{appGFD} or Ref.\ \cite{WALBORN201087} for definitions of the coefficients.

Besides the impulse response function, an optical system can also be expressed as a quantum operator acting on the photon transverse spatial state, and some of the following expressions can be simplified (see Appendix \ref{OSaQO}).

\subsection{Description of quantum monochromatic optical fields}

A single-photon state with a certain polarization in the free space is denoted by $|\Psi\rangle=\int d\mathbf{k}\tilde{\psi}(\mathbf{k})\hat{a}_\mathbf{k}^\dagger|\mathrm{vac}\rangle$, where $|\mathrm{vac}\rangle$ is the vacuum state. For a monochromatic light field, the $\sqrt{\omega(\mathbf{k})}=\sqrt{c|\mathbf{k}|}$ term in the electric field operator becomes a constant, and the photon wave function in the position basis can be defined as $\psi(\mathbf{r})=\langle\mathrm{vac}|\hat{E}^{(+)}(\mathbf{r})|\Psi\rangle$ \cite{shihlzgx}, where the positive electric field operator $\hat{E}^{(+)}(\mathbf{r})\propto\hat{a}_{\mathbf{r}}$, and it is the inverse Fourier transform of $\tilde{\psi}(\mathbf{k})$.

Then, we ignore the normalization, do not distinguish $\hat{E}^{(+)}(\mathbf{r})$ and $\hat{a}_\mathbf{r}$, and consider a general space with linear, passive media (e.g., an optical system without light sources). The state $\int d\mathbf{r}\psi(\mathbf{r})\hat{a}_\mathbf{r}^\dagger|\mathrm{vac}\rangle$ with an arbitrary $\psi(\mathbf{r})$ cannot guarantee the optical field is monochromatic, so, with the three-dimensional impulse response function $h(\mathbf{r},\mathbf{r}_0)$ which describes the propagated electric field amplitude form a monochromatic point source at $\mathbf{r}$, we define the monochromatic creation operator
\begin{equation}
    \hat{a}_{k,\mathbf{r}}^\dagger=\int d\mathbf{r}'h(\mathbf{r}',\mathbf{r})\hat{a}_{\mathbf{r}'}^\dagger.
\end{equation}
For such a state $|\Psi\rangle=\int d\mathbf{r}\psi_0(\mathbf{r})\hat{a}_{k,\mathbf{r}}^\dagger|\mathrm{vac}\rangle$, the resultant photon wave function is not $\psi_0(\mathbf{r})$ (which we name the \emph{creation amplitude}), but
\begin{equation}
    \psi(\mathbf{r})=\langle\mathrm{vac}|\hat{a}_\mathbf{r}|\Psi\rangle=\int d\mathbf{r}_0\psi_0(\mathbf{r}_0)h(\mathbf{r},\mathbf{r}_0).
\end{equation}
Notice that if $\mathbf{r}$ is inside a medium, the classical intensity is also proportional to $n(\mathbf{r})$ while $|\psi(\mathbf{r})|^2$ is not, so we avoid the discussion of photon detection inside a medium.

In the free space or an infinite LHI medium, the propagated field from a point is a spherical wave. Under the paraxial approximation, we have $\hat{a}_{k,\mathbf{r}}^\dagger=\int d\boldsymbol{\rho}'dz'G_{|z'|}(\boldsymbol{\rho}')\hat{a}_{\boldsymbol{\rho}+\boldsymbol{\rho}',z+z'}^\dagger$, where $G_{|z|}(\boldsymbol{\rho})$ describes a paraxial spherical wave traveling both forward and backward \cite{gezabs}, while $G_z(\boldsymbol{\rho})$ is a wave traveling forward which is focused at $\mathbf{0}$.

A distinguishable two-photon field can be photons with different wavelengths or photons with the same wavelength but orthogonal polarizations. Taking the first case as an example, with the state $|\Psi\rangle=\int d\mathbf{r}_1d\mathbf{r}_2\psi_0(\mathbf{r}_1,\mathbf{r}_2)\hat{a}_{k_1,\mathbf{r}_1}^\dagger\hat{a}_{k_2,\mathbf{r}_2}^\dagger|\mathrm{vac}\rangle$ and the impulse response functions for the two wavelengths $h_1(\mathbf{r},\mathbf{r}_0)$ and $h_2(\mathbf{r},\mathbf{r}_0)$, their wave function
\begin{align}\label{distwf}
    &\psi(\mathbf{r}_1,\mathbf{r}_2)=\langle\mathrm{vac}|\hat{a}_{\mathbf{r}_1}\hat{a}_{\mathbf{r}_2}|\Psi\rangle\nonumber\\
    =&\int d\mathbf{r}_{10}d\mathbf{r}_{20}\psi_0(\mathbf{r}_{10},\mathbf{r}_{20})h_1(\mathbf{r}_1,\mathbf{r}_{10})h_2(\mathbf{r}_2,\mathbf{r}_{20}),
\end{align}
where $\hat{a}_{\mathbf{r}_1}$ and $\hat{a}_{\mathbf{r}_2}$ only annihilate the fields with $k_1$ and $k_2$, respectively (after filtering out unwanted frequencies, they are ordinary annihilation operators), and the JPD is $|\psi(\mathbf{r}_1,\mathbf{r}_2)|^2$. If they are indistinguishable, we use the same impulse response function, and the result
\begin{align}\label{indistwf}
    \int&d\mathbf{r}_{10}d\mathbf{r}_{20}\psi_0(\mathbf{r}_{10},\mathbf{r}_{20})\nonumber\\
    &\times[h(\mathbf{r}_1,\mathbf{r}_{10})h(\mathbf{r}_2,\mathbf{r}_{20})+h(\mathbf{r}_2,\mathbf{r}_{10})h(\mathbf{r}_1,\mathbf{r}_{20})]
\end{align}
has exchange symmetry. If $\psi_0(\mathbf{r}_1,\mathbf{r}_2)$ is symmetric itself, Eq.\ \eqref{indistwf} can be simplified to Eq.\ \eqref{distwf}.

\subsection{Optical reciprocity}

The reciprocity theorem \cite{nanooptics} is that in a space with linear, passive media (which can be inhomogeneous, anisotropic, or absorptive), considering two sets of electric current distributions oscillating at the same frequency, letting their complex amplitudes be $\mathbf{J}_1(\mathbf{r})$ and $\mathbf{J}_2(\mathbf{r})$, and their individually produced electric field complex amplitudes be $\mathbf{E}_1(\mathbf{r})$ and $\mathbf{E}_2(\mathbf{r})$, respectively, for a region $V$ covering all the sources,
\begin{equation}\label{rcrt}
    \int_V d\mathbf{r}[\mathbf{E}_1(\mathbf{r})\cdot\mathbf{J}_2(\mathbf{r})-\mathbf{E}_2(\mathbf{r})\cdot\mathbf{J}_1(\mathbf{r})]=0.
\end{equation}
If the two sources are point-like $\mathbf{J}_1(\mathbf{r})=\mathbf{J}_1\delta(\mathbf{r}-\mathbf{r}_1)$ and $\mathbf{J}_2(\mathbf{r})=\mathbf{J}_2\delta(\mathbf{r}-\mathbf{r}_2)$, we have $\mathbf{E}_1(\mathbf{r}_2)\cdot\mathbf{J}_2=\mathbf{E}_2(\mathbf{r}_1)\cdot\mathbf{J}_1$. This theorem can be broken by, for example, the Faraday effect. Considering an optical system insensitive to polarization, denoting the inverse optical system of $h(\boldsymbol{\rho},\boldsymbol{\rho}_0)$ as $\bar{h}(\boldsymbol{\rho},\boldsymbol{\rho}_0)$, we have $\bar{h}(\boldsymbol{\rho},\boldsymbol{\rho}_0)=h(\boldsymbol{\rho}_0,\boldsymbol{\rho})$. The bar is unnecessary in three-dimensional functions $h(\mathbf{r},\mathbf{r}_0)=h(\mathbf{r}_0,\mathbf{r})$. 

Then, we arrange horizontal ($H$) and vertical ($V$) directions for $\mathbf{r}_1$ and $\mathbf{r}_2$ respectively (the coordinate axes are often different in an optical system with reflection or refraction). When the source intensities are equal, denoting a certain polarization $\sigma$ as $H$ or $V$, then $U_{\sigma_2}(\mathbf{r}_2)$ produced by a $\sigma_1$ source at $\mathbf{r}_1$ equals $U_{\sigma_1}(\mathbf{r}_1)$ produced by a $\sigma_2$ source at $\mathbf{r}_2$. Denoting the impulse response function of $\sigma$ polarization from a $\sigma_0$ source as $h_{\sigma,\sigma_0}(\mathbf{r},\mathbf{r}_0)$, then $h_{\sigma,\sigma_0}(\mathbf{r},\mathbf{r}_0)=h_{\sigma_0,\sigma}(\mathbf{r}_0,\mathbf{r})$.

\subsection{Spontaneous parametric down-conversion}

When considering a specific group of polarizations of the pump (\emph{p}), signal (\emph{s}), and idler (\emph{i}) light, the nonlinear part of the Hamiltonian inside an NLC is
\begin{equation}
    \hat{H}=\varepsilon_0\int d\mathbf{r}\chi^{(2)}(\mathbf{r})\hat{E}_p(\mathbf{r})\hat{E}_s(\mathbf{r})\hat{E}_i(\mathbf{r}),
\end{equation}
where the second-order susceptibility is real and generally position-dependent (it is zero outside the crystal). We consider perfect energy conservation (steady state), the first-order approximation (low-gain), and the $\hat{E}_p^{(+)}(\mathbf{r})\hat{E}_s^{(-)}(\mathbf{r})\hat{E}_i^{(-)}(\mathbf{r})$ term \cite{WALBORN201087,Schneeloch_2016,Karan_2020}. The pump beam is often a monochromatic classical laser, which is described by a multimode coherent state in quantum optics. From the first-order susceptibility (the refractive index and linear absorptive coefficient) distribution of the space, its complex amplitude $U_p(\mathbf{r})$ can be determined. Denoting the wave numbers of the \emph{s} and \emph{i} light in the vacuum as $k_s$ and $k_i$, the biphoton state is
\begin{equation}\label{spdcstate}
    |\Psi\rangle=\int d\mathbf{r}\chi^{(2)}(\mathbf{r})U_p(\mathbf{r})\hat{a}_{k_s,\mathbf{r}}^\dagger\hat{a}_{k_i,\mathbf{r}}^\dagger|\mathrm{vac}\rangle.
\end{equation}
So, the photon pair is created at the same position with the amplitude $\chi^{(2)}(\mathbf{r})U_p(\mathbf{r})$, and
\begin{equation}
    \psi_0(\mathbf{r}_1,\mathbf{r}_2)=\chi^{(2)}(\mathbf{r}_1)U_p(\mathbf{r}_1)\delta(\mathbf{r}_1-\mathbf{r}_2)
\end{equation}
is naturally symmetric.

In the standard approach, the following analysis of the biphoton state is in the momentum space, where the angular spectrum of the pump beam affects the momentum anticorrelation strength and the phase matching condition limits the width of momentum distribution (Appendix \ref{SPDCAMS}). In Sec.\ \ref{aspdcbps}, we will use the AWP to analyze the SPDC state in the position basis.

\subsection{Monochromatic advanced-wave picture}

\subsubsection{Essential case}

\begin{figure}[t]
\includegraphics[width=0.48\textwidth]{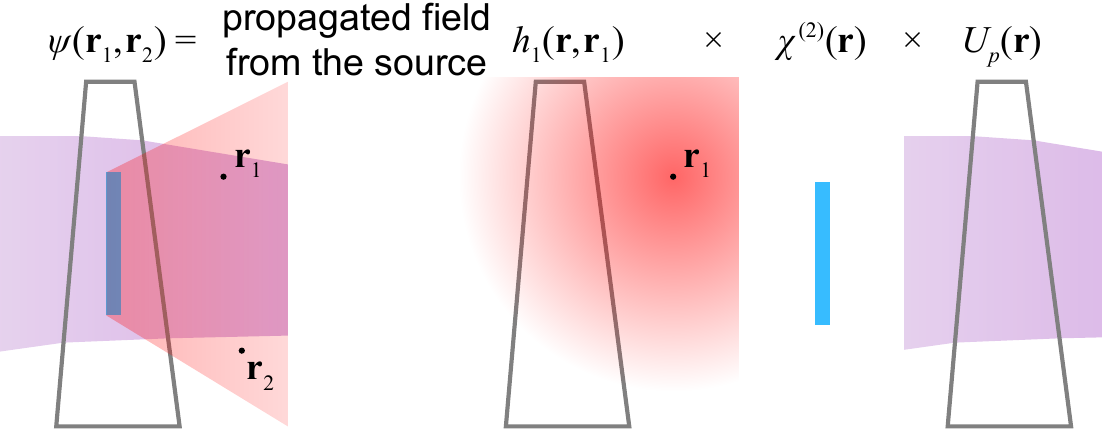}
\caption{\label{principlefig}The principle of monochromatic AWP. The biphoton wave function $\psi(\mathbf{r}_1,\mathbf{r}_2)$ from SPDC is calculated by the propagated field from a source with a creation amplitude distribution which is the optical field amplitude from a point source at $\mathbf{r}_1$ times the second-order susceptibility distribution and the pump beam amplitude.}
\end{figure}

After biphoton creation, according to Eq.\ \eqref{distwf},
\begin{align}\label{awpcore}
    \psi(\mathbf{r}_1,\mathbf{r}_2)=&\int d\mathbf{r}_0\chi^{(2)}(\mathbf{r}_0)U_p(\mathbf{r}_0)h_1(\mathbf{r}_1,\mathbf{r}_0)h_2(\mathbf{r}_2,\mathbf{r}_0)\nonumber\\
    =&\int d\mathbf{r}h_1(\mathbf{r},\mathbf{r}_1)\chi^{(2)}(\mathbf{r})U_p(\mathbf{r})h_2(\mathbf{r}_2,\mathbf{r})\nonumber\\
    =&\int d\mathbf{r}h_2(\mathbf{r},\mathbf{r}_2)\chi^{(2)}(\mathbf{r})U_p(\mathbf{r})h_1(\mathbf{r}_1,\mathbf{r}).
\end{align}
Here, the down-converted photons undergo linear processes after creation, so $h_1(\mathbf{r},\mathbf{r}_0)$ and $h_2(\mathbf{r},\mathbf{r}_0)$ are calculated from the first-order susceptibilities of the space as usual. From the second line of Eq.\ \eqref{awpcore}, the biphoton wave function can be understood in this way: the complex amplitude of the light emitting from $\mathbf{r}_1$ with the wave number $k_1$ (the advanced wave) is multiplied by $\chi^{(2)}(\mathbf{r})U_p(\mathbf{r})$, the product is the new photon creation amplitude with the wave number $k_2$ (and a new polarization in type-II SPDC), and the wave function is the amplitude of the new field propagated to $\mathbf{r}_2$, as shown in Fig.\ \ref{principlefig}. The indices $1$ and $2$ can be interchanged, as shown in the last line of Eq.\ \eqref{awpcore}. This is the main idea of the AWP for monochromatic pump, signal, and idler lights. Aichele \emph{et al.}\ suggested the multiplication in creation amplitude calculation can be considered as a difference frequency generation with the pump beam and the AWP light field from $\mathbf{r}_1$ which is reversed again (propagating forward from the back of the crystal) \cite{Aichele2002}. Alternatively, it can be modeled it as a sum-frequency generation from both lights propagating backward \cite{PhysRevLett.117.123901,Lenzini2018} (also known as the ``double Klyshko picture'' \cite{PhysRevA.80.053820}).

The better-known AWP is a special case where the crystal is a sufficiently thin plane with a constant $\chi^{(2)}$ and $\psi(\boldsymbol{\rho}_1,\boldsymbol{\rho}_2)=\int d\boldsymbol{\rho}\bar{h}_1(\boldsymbol{\rho},\boldsymbol{\rho}_1)U_p(\boldsymbol{\rho})h_2(\boldsymbol{\rho}_2,\boldsymbol{\rho})$, so the crystal becomes a mirror with the reflectivity $U_p(\boldsymbol{\rho})$. It does not take the crystal thickness into account (the two photons are perfectly correlated in position), so, for some quantum imaging protocols, properties like the spatial resolution cannot be analyzed.

In GI, one photon is not postselected to a point, but a surface region via a bucket detector which triggers the detector of the other photon. In QIUP, one photon is not detected at all. These two situations should be addressed before analyzing their setups, as shown in Appendix \ref{AWPothercases}, where AWP theories of partially coherent pump beams and $N$ photons \cite{PhysRevA.76.045802} are also presented. The result is this: in the case of bucket detection, the whole surface of the bucket detector emits advanced waves incoherently; in the case of no detection, if the down-converted photons either travel to infinity (without the detectors) or get absorbed, a surface covering the whole setup and regions with absorption both emit advanced waves incoherently.

\subsubsection{Case of pure-state postselection and the linearity in AWP}\label{TCGSPS}

Sometimes, one photon is postselected to a pure state (for example, being collected by a single-mode fiber \cite{Mair2001}). Letting the postselected wave function defined on a certain plane $S_1$ be $\psi_1(\mathbf{r})$, the annihilation operator $\hat{a}_{\mathbf{r}_1}$ is replaced by $\big[\int_{S_1}d\mathbf{r}\psi_1(\mathbf{r})\hat{a}_{\mathbf{r}}^\dagger\big]^\dagger=\int_{S_1}d\boldsymbol{\mathbf{r}}\psi_1^\ast(\mathbf{r})\hat{a}_{\mathbf{r}}$, and we can superpose Eq.\ \eqref{awpcore} over different $\mathbf{r}_1$ values
\begin{align}\label{generalps}
    \psi(\mathbf{r}_2|\psi_1)&=\int d\mathbf{r}\left[\int_{S_1}d\mathbf{r}_1\psi_1^\ast(\mathbf{r}_1)h_1(\mathbf{r},\mathbf{r}_1)\right]\nonumber\\
    &\qquad\quad\ \times\chi^{(2)}(\mathbf{r})U_p(\mathbf{r})h_2(\mathbf{r}_2,\mathbf{r})\nonumber\\
    &=\int_{S_1}d\mathbf{r}_1\psi_1^\ast(\mathbf{r}_1)\psi(\mathbf{r}_1,\mathbf{r}_2).
\end{align}
The first line means that in the AWP, before multiplying $\chi^{(2)}(\mathbf{r})U_p(\mathbf{r})$, the optical field (the expression in the brackets) is now from the creation amplitude $\psi_1^\ast(\mathbf{r})$ (the postselected state traveling backward) rather than a point unless $\psi_1^\ast(\mathbf{r})=\delta(\mathbf{r}-\mathbf{r}_1)$. We can also imagine that after $h_1(\mathbf{r},\mathbf{r}_0)$, a hypothetical optical system $h_\mathrm{ps}(\mathbf{r},\mathbf{r}_0)=\psi_1^\ast(\mathbf{r}_0)\delta(\mathbf{r}-\mathbf{r}_\mathrm{ps})$ maps the state $\psi_1(\mathbf{r})$ to the point $\mathbf{r}_\mathrm{ps}$ at a plane $S_\mathrm{ps}$ behind $S_1$ and absorbs any state components orthogonal to it. The expression of $h_\mathrm{ps}(\mathbf{r},\mathbf{r}_0)$ is applicable only when $\mathbf{r}$ is at $S_\mathrm{ps}$ and $\mathbf{r}_0$ is at $S_1$, so the inverse form $\bar{h}_\mathrm{ps}(\mathbf{r},\mathbf{r}_0)=\psi_1^\ast(\mathbf{r})\delta(\mathbf{r}_0-\mathbf{r}_\mathrm{ps})$ \cite{rhoorr}. The inverse of the combined system for photon 1 is
\begin{equation}
    \bar{h}_{1\mathrm{ps}}(\mathbf{r},\mathbf{r}_0)=\int_{S_1}d\mathbf{r}_1h_1(\mathbf{r},\mathbf{r}_1)\bar{h}_\mathrm{ps}(\mathbf{r}_1,\mathbf{r}_0).
\end{equation}
Substituting $h_1(\mathbf{r},\mathbf{r}_1)$ with $\bar{h}_{1\mathrm{ps}}(\mathbf{r},\mathbf{r}_\mathrm{ps})$ in the second line of Eq.\ \eqref{awpcore} yields Eq.\ \eqref{generalps}.

Note that at $S_1$, the input field traveling backward is $\psi_1^\ast(\mathbf{r})$, and from the second line of Eq.\ \eqref{generalps}, the photon 2 wave function can be interpreted as the field after a classical linear optical setup whose impulse response function $h(\mathbf{r},\mathbf{r}_0)=\psi(\mathbf{r}_0,\mathbf{r})$. If the two photons are indistinguishable, the reciprocity $h(\mathbf{r},\mathbf{r}_0)=h(\mathbf{r}_0,\mathbf{r})$ is satisfied. In Sec.\ \ref{aspdcbps}, we will derive the biphoton wave functions from SPDC, and present possible realizations of the corresponding classical processes in the AWP-unfolded setup.

\subsubsection{Case including polarization}\label{TCIP}

If different polarizations are considered (for example, using two NLCs with different optical axis directions can produce polarization-entangled biphotons \cite{Defienne2021}), the pump beam is described by $U_{p,H}(\mathbf{r})$ and $U_{p,V}(\mathbf{r})$. Letting $\sigma=H,V$ be the index of polarization, the biphoton state is
\begin{equation} 
    |\Psi\rangle=\!\sum_{\sigma_p,\sigma_1,\sigma_2}\int d\mathbf{r}\chi_{\sigma_p\sigma_1\sigma_2}^{(2)}(\mathbf{r})U_{p,\sigma_p}(\mathbf{r})\hat{a}_{\sigma_1,k_1,\mathbf{r}}^\dagger\hat{a}_{\sigma_2,k_2,\mathbf{r}}^\dagger|\mathrm{vac}\rangle,
\end{equation}
where $\hat{a}_{\sigma,k,\mathbf{r}}^\dagger=\sum_{\sigma'}\int d\mathbf{r}'h_{\sigma',\sigma}(\mathbf{r}',\mathbf{r})\hat{a}_{\sigma',\mathbf{r}'}^\dagger$. The postselected photon 1 state is at $\mathbf{r}_1$ with a specific polarization $P=(c_H,c_V)$ ($|c_H|^2+|c_V|^2=1$), described by the creation operator $\hat{a}_{P,\mathbf{r}_1}^\dagger=c_H\hat{a}_{H,\mathbf{r}_1}^\dagger+c_V\hat{a}_{V,\mathbf{r}_1}^\dagger$. Denoting the conjugate polarization as $P^\ast=(c_H^\ast,c_V^\ast)$ (invariant for linear polarizations) and the impulse response function from a source with $P^\ast$ polarization as $h_{\sigma,P^\ast}(\mathbf{r},\mathbf{r}_0)=\sum_{\sigma_0}c_{\sigma_0}^\ast h_{\sigma,\sigma_0}(\mathbf{r},\mathbf{r}_0)$, the conditional wave function of photon 2 with $\sigma_2$ polarization is
\begin{align}
    \langle\mathrm{vac}|\hat{a}_{P,\mathbf{r}_1}\hat{a}_{\sigma_2,\mathbf{r}_2}|\Psi\rangle=&\sum_{\sigma_p,\sigma'_1,\sigma'_2}\int d\mathbf{r}h_{\sigma'_1,P^\ast}(\mathbf{r},\mathbf{r}_0)\nonumber\\
    &\ \ \times\chi_{\sigma_p\sigma'_1\sigma'_2}^{(2)}(\mathbf{r})U_{p,\sigma_p}(\mathbf{r})h_{\sigma_2,\sigma'_2}(\mathbf{r}_2,\mathbf{r}),
\end{align}
which means the $\sigma'_1$ component of the field created by the source with $P^\ast$ polarization is multiplied by $\sum_{\sigma_p}\chi_{\sigma_p\sigma'_1\sigma'_2}^{(2)}(\mathbf{r})U_{p,\sigma_p}(\mathbf{r})$ as the new creation amplitude of $\sigma'_2$ polarization, and the result is the $\sigma_2$ component of the new field at $\mathbf{r}_2$ created by the two $\sigma'_2$ polarizations. In Sec.\ \ref{QHwPEsubsect}, we will use the AWP with polarization to analyze quantum holography with polarization entanglement \cite{Defienne2021}.

\section{Analysis of SPDC biphoton state}\label{aspdcbps}

\subsection{Degenerate collinear type-I SPDC}\label{dcispdcsub}

In this section, we use the AWP to analyze biphoton position wave functions of degenerate ($k_p=2k_s=2k_i$, which we denote by $2k$) collinear type-I ($e\to o+o$ for a negative crystal) SPDC pumped by a normally incident plane wave, where the effective refractive index $\eta$ (see Appendix \ref{appGFD}) of the pump beam equals $n_o$ of the down-converted photons. An NLC with a constant $\chi^{(2)}$, an infinite transverse size, and the thickness $L$ (which we name a bulk crystal) is placed at $-L/2<z<L/2$, so the pump beam amplitude $U_p(\mathbf{r})=e^{i2n_o kz}$ inside the crystal. From the standard analysis in the momentum space, the biphoton wave function at $z=0$ is
\begin{equation}\label{standardmspdccol}
    \tilde{\psi}(\mathbf{q}_1,\mathbf{q}_2)=\delta(\mathbf{q}_1+\mathbf{q}_2)\operatorname{sinc}\left(\frac{L}{8n_ok}|\mathbf{q}_1-\mathbf{q}_2|^2\right),
\end{equation}
where $\operatorname{sinc}(x)=\sin x/x$ and $\operatorname{sinc}(0)=1$. By inverse Fourier transform, the position wave function
\begin{equation}\label{standardpspdccol}
    \psi(\boldsymbol{\rho}_1,\boldsymbol{\rho}_2)=\operatorname{Ssi}\left(\frac{n_ok|\boldsymbol{\rho}_1-\boldsymbol{\rho}_2|^2}{2L}\right),
\end{equation}
where $\operatorname{Ssi}(x)=\int_0^x\operatorname{sinc}(x')dx'-\pi/2$ is the shifted sine integral function \cite{PhysRevA.95.063836}. If $z\neq0$, from the angular spectrum theory, a paraboloid phase is added to Eq.\ \eqref{standardmspdccol}, but the position wave function has a more complicated expression \cite{PhysRevA.110.033713} unless the Gaussian approximation is used (see Sec.\ \ref{PdGs}).

Here, we derive Eq.\ \eqref{standardpspdccol} using the AWP. However, if we choose the points $\mathbf{r}_1$ and $\mathbf{r}_2$ at the middle plane of the crystal $z=0$, the biphoton created in the region $0<z<L/2$ propagates backward to the two positions in the standard approach, and the result is unwanted. Instead, we should let $\mathbf{r}_1$ and $\mathbf{r}_2$ be at the back focal plane $z=z_b$ of a combination of two perfect $4f$ systems with unit magnification ratio \cite{8fsystem}. Only the advanced wave traveling \emph{backward} from $\mathbf{r}_1$ arrives at the crystal, and the waves traveling \emph{forward} created in the crystal contribute to the result. Letting the forward direction be $\mathbf{e}_z$, we write $\hat{a}_{k\pm,\mathbf{r}}^\dagger=\int d\mathbf{r}'h_\pm(\mathbf{r}',\mathbf{r})\hat{a}_{\mathbf{r}'}^\dagger$, and the reciprocity theorem is $h_+(\mathbf{r},\mathbf{r}_0)=h_-(\mathbf{r}_0,\mathbf{r})$. Because of the perfect imaging, when $\mathbf{r}$ is inside the crystal, $h(\mathbf{r},(\boldsymbol{\rho}_0,z_b))=h_-(\mathbf{r},(\boldsymbol{\rho}_0,z_b))=h_-(\mathbf{r},(\boldsymbol{\rho}_0,0))$; when $\mathbf{r}_0$ is inside the crystal, $h((\boldsymbol{\rho},z_b),\mathbf{r}_0)=h_+((\boldsymbol{\rho},z_b),\mathbf{r}_0)=h_+((\boldsymbol{\rho},0),\mathbf{r}_0)$. Under the paraxial approximation (which is also used in the standard momentum analysis), if $\mathbf{r}$ and $\mathbf{r}_0$ are both inside the crystal, $h_\pm(\mathbf{r},\mathbf{r}_0)=G_{\pm(z-z_0)}(\boldsymbol{\rho}-\boldsymbol{\rho}_0)$ (the refractive index is $n_o$). From a modified Eq.\ \eqref{spdcstate},
\begin{equation}\label{spdcstatedci}
    |\Psi\rangle=\int d\mathbf{r}\chi^{(2)}(\mathbf{r}_0)U_p(\mathbf{r}_0)\hat{a}_{k+,\mathbf{r}}^\dagger\hat{a}_{k+,\mathbf{r}}^\dagger|\mathrm{vac}\rangle,
\end{equation}
and the wave function \cite{replace12}
\begin{align}\label{distwfforward}
    \psi(\mathbf{r}_1,\mathbf{r}_2)&=\langle\mathrm{vac}|\hat{a}_{\mathbf{r}_1}\hat{a}_{\mathbf{r}_2}|\Psi\rangle\nonumber\\
    &=\int d\mathbf{r}h_-(\mathbf{r},\mathbf{r}_1)\chi^{(2)}(\mathbf{r})U_p(\mathbf{r})h_+(\mathbf{r}_2,\mathbf{r}).
\end{align}
$z_1$ and $z_2$ which should be $z_b$ can be set to $0$, and
\begin{align}\label{distwfdci}
    \psi(\boldsymbol{\rho}_1,\boldsymbol{\rho}_2)&=\int_\textrm{crystal}d\mathbf{r}G_{-z}(\boldsymbol{\rho}-\boldsymbol{\rho}_1)e^{i2n_o kz}G_{-z}(\boldsymbol{\rho}_2-\boldsymbol{\rho})\nonumber\\
    &\propto\int d\boldsymbol{\rho}\int_{-L/2}^{L/2}\frac{dz}{z^2}e^{-in_ok(|\boldsymbol{\rho}-\boldsymbol{\rho}_1|^2+|\boldsymbol{\rho}_2-\boldsymbol{\rho}|^2)/(2z)}\nonumber\\
    &\propto\operatorname{Ssi}\left(\frac{n_ok|\boldsymbol{\rho}_1-\boldsymbol{\rho}_2|^2}{2L}\right),
\end{align}
which is the same as Eq.\ \eqref{standardpspdccol} (see Appendix \ref{CalcInt} for the detailed calculation). The calculation of degenerate beamlike type-II SPDC \cite{Takeuchi:01} is shown in Appendix \ref{CDBLT2SPDC}, where a tilt phase is present. Although the calculation in the position space is more complicated, and the first line of Eq.\ \eqref{distwfdci} also results from the standard approach in Eq.\ \eqref{spdcstate} \cite{PhysRevA.80.053820}, the AWP provides a classical-like way to deduce the form of biphoton wave function.

Note that in the second line of Eq.\ \eqref{distwfdci}, a term like $e^{iaz}$ has been eliminated because of the collinear phase matching. If the two points are at the back of the crystal $z_1,z_2<-L/2$, in the AWP, the initial wave from $\mathbf{r}_1$ propagating forward has a phase like $e^{in_okz}$ (omitting the transverse phase difference) in the crystal, which becomes the creation amplitude $e^{i3n_okz}$ after multiplying the pump light amplitude. Compared to the previous case $e^{in_okz}$ (a realistic wave propagating forward), the difference is $e^{i2n_okz}$, which is highly destructive after integrating over $z$ inside the crystal, so SPDC photons are very unlikely to travel backward. If the phase matching cannot be satisfied, i.e., $\eta\neq n_o$ with a different optical axis angle, the difference between the creation amplitude and the forward propagating wave is $e^{i2(\eta-n_o)kz}$, which enables noncollinear phase matching if $\eta<n_o$, but the wave function calculation is more difficult. If the pump beam is an oblique plane wave, $k_z\neq\eta k$ inside the crystal, and the phase matching may be realized. This corresponds to tilting the crystal, a common experimental method to adjust the phase matching. Quasiphase matching means $\chi^{(2)}(\mathbf{r})$ is modulated to $C\operatorname{sgn}\{\cos[2(\eta-n_o)kz]\}$, which prevents the destructive integration to some extent. Besides modulating the pump field, if we want to add a phase $e^{i\Phi(\mathbf{r})}$ to the creation amplitude, $\chi^{(2)}(\mathbf{r})$ can be designed as $C\operatorname{sgn}[\cos\Phi(\mathbf{r})]$. Like holography, a conjugate term $e^{-i\Phi(\mathbf{r})}$ will also be present.

\subsection{AWP-unfolded setup of an SPDC crystal}\label{AWPESSPDCC}

\begin{figure}[t]
\includegraphics[width=0.48\textwidth]{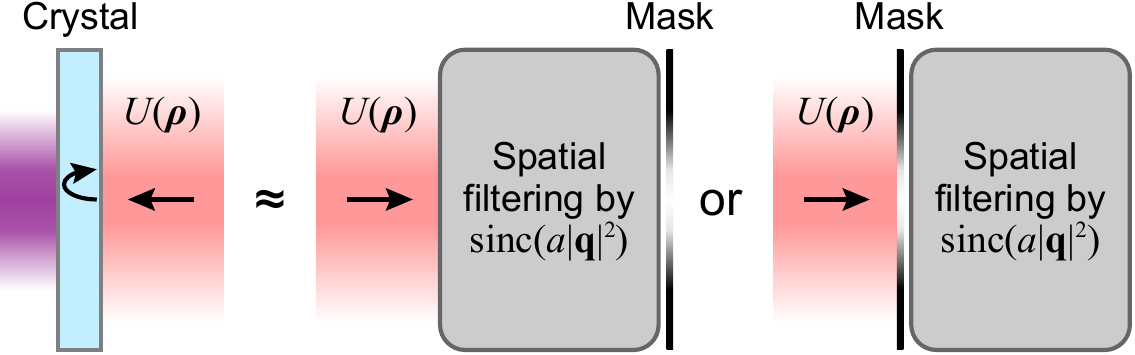}
\caption{\label{equivalfig}In the AWP-unfolded setup, if the pump beam walk-off can be ignored, a bulk collinear type-I SPDC crystal is approximately equivalent to a spatial (angular spectrum) filtering system and a mask determined by the pump beam amplitude. The filtering system and the mask can be exchanged.}
\end{figure}

A bulk SPDC crystal can become a familiar optical device in the AWP-unfolded setup. From Sec.\ \ref{TCGSPS}, considering the hypothetical two $4f$ systems in Sec.\ \ref{dcispdcsub}, a point source at $\boldsymbol{\rho}_1$ of the back focal plane traveling backward becomes a light field traveling forward with the amplitude $\psi(\boldsymbol{\rho}_1,\boldsymbol{\rho})$. The polarization or the wavelength may be changed in type-II or nondegenerate SPDC. Nevertheless, if we solely care about the transverse spatial amplitude, in the AWP, denoting the incoming field propagating backward as $\psi_\textrm{in}(\boldsymbol{\rho}_1)$, the wave function of the reflected photon is
\begin{equation}\label{crystreflect}
    \psi(\boldsymbol{\rho}_2)=\int d\boldsymbol{\rho}_1\psi_\textrm{in}(\boldsymbol{\rho}_1)\psi(\boldsymbol{\rho}_1,\boldsymbol{\rho}_2).
\end{equation}
In degenerate collinear type-I SPDC, Eq.\ \eqref{standardpspdccol} or \eqref{distwfdci} is a function of $\boldsymbol{\rho}_1-\boldsymbol{\rho}_2$, so Eq.\ \eqref{crystreflect} has the form of convolution. Therefore, if the pump beam is a normally incident plane wave, the NLC is a spatial filtering system [the coherence transfer function has the form of $\operatorname{sinc}(a|\mathbf{q}|^2)$]. If an arbitrary pump beam whose width is much larger than $\alpha L$ is used (ignoring its walk-off), the form of the impulse response function is Eq.\ \eqref{colapproxspdcwf}. When the correlation (Ssi) term is much narrower than the pump term, an approximation $\boldsymbol{\rho}_1+\boldsymbol{\rho}_2\approx2\boldsymbol{\rho}_2$ can be made, and
\begin{equation}\label{colapproxspdcwfapprox}
    \psi(\boldsymbol{\rho}_1,\boldsymbol{\rho}_2)=\operatorname{Ssi}\left(\frac{n_ok|\boldsymbol{\rho}_1-\boldsymbol{\rho}_2|^2}{2L}\right)U_p(\boldsymbol{\rho}_2),
\end{equation}
which means the unfolded setup is a convolution with $\operatorname{Ssi}[n_ok|\boldsymbol{\rho}|^2/(2L)]$ followed by amplitude modulation $U_p(\boldsymbol{\rho})$. In the nondegenerate case, borrowed from Eq.\ \eqref{colapproxspdcmwfnd}, Eq.\ \eqref{colapproxspdcwfapprox} becomes
\begin{equation}\label{colapproxspdcwfapproxndg}
    \psi(\boldsymbol{\rho}_1,\boldsymbol{\rho}_2)=\operatorname{Ssi}\left(\frac{2\pi|\boldsymbol{\rho}_1-\boldsymbol{\rho}_2|^2}{L\lambda_+}\right)U_p(\boldsymbol{\rho}_2),
\end{equation}
where $\lambda_+=\lambda_s/n_s+\lambda_i/n_i$ is the sum of the two wavelengths inside the crystal. An alternative approximation $\boldsymbol{\rho}_1+\boldsymbol{\rho}_2\approx2\boldsymbol{\rho}_1$ can also be used if $U_p(\boldsymbol{\rho})$ is roughly unchanged after the convolution by the Ssi function, and the result is performing convolution after amplitude modulation, i.e., replacing $U_p(\boldsymbol{\rho}_2)$ in Eqs.\ \eqref{colapproxspdcwfapprox} and \eqref{colapproxspdcwfapproxndg} by $U_p(\boldsymbol{\rho}_1)$. The two approximations are illustrated in Fig.\ \ref{equivalfig}.

Numerically, the full width at half maximum (FWHM) of the point spread function (PSF, the squared modulus of the impulse response function) is $\Delta\rho_\mathrm{d}\approx0.770\sqrt{L\lambda/n_o}$ for the degenerate case or $\Delta\rho_\mathrm{nd}\approx0.544\sqrt{L\lambda_+}$ for the nondegenerate case. The FWHM of the squared coherence transfer function is $\Delta q_\mathrm{d}\approx11.83/\sqrt{L\lambda/n_o}$ (degenerate) or $\Delta q_\mathrm{nd}\approx16.73/\sqrt{L\lambda_+}$ (nondegenerate). These data can be used to analyze the spatial resolution and field of view of GI and QIUP. 

\subsection{Propagation of the double-Gaussian approximated biphoton state}\label{PdGs}

To study the propagation of the biphoton created by the SPDC, the Ssi term in the position wave function or the sinc term in the momentum one is often approximated by a Gaussian function. A possible approximation is $\operatorname{sinc}(ax^2)\approx\exp(-0.455ax^2)$ such that the two squared functions reach $1/e^2$ of the maximum ($x=0$) at the same $x$ \cite{PhysRevA.75.050101} (the authors of Ref.\ \cite{PhysRevA.106.063714} discussed other coefficients and functions). If the pump beam is also Gaussian, the biphoton state is known as the double-Gaussian state \cite{PhysRevLett.92.127903}
\begin{equation}\label{dGwf}
    \psi(\boldsymbol{\rho}_1,\boldsymbol{\rho}_2)=\exp\left(-\frac{|\boldsymbol{\rho}_1+\boldsymbol{\rho}_2|^2}{4\sigma_+^2}-\frac{|\boldsymbol{\rho}_1-\boldsymbol{\rho}_2|^2}{4\sigma_-^2}\right),
\end{equation}
whose free-space propagation can be easily calculated from the angular spectrum theory ($\sigma_\pm^2\to\sigma_\pm^2+iz/k$ after propagating the distance $z$) \cite{PhysRevA.75.050101,PhysRevA.95.063836,Schneeloch_2016}. Our group measured the propagated biphoton wave functions using quantum Shack-Hartmann wavefront sensing \cite{qshws}.

\begin{figure}[t]
\includegraphics[width=0.48\textwidth]{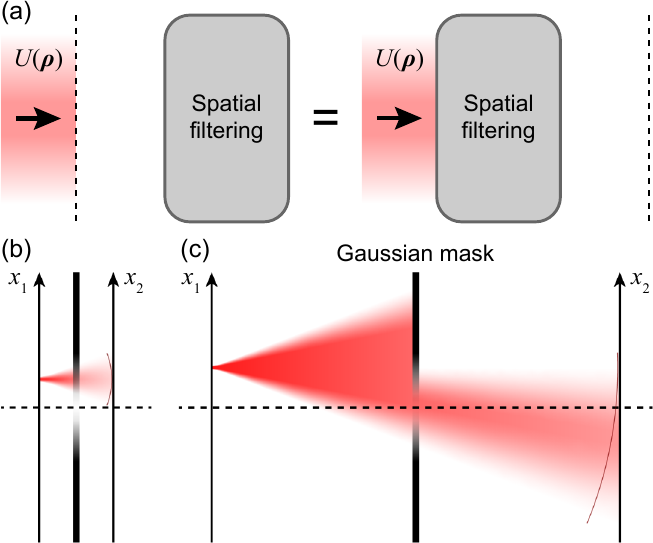}
\caption{\label{dgproplfig}(a) In classical optics, free-space propagation is a convolution process, and can be interchanged with a spatial filtering system. (b) The unfolded setup of a double-Gaussian state propagated by a shorter distance. The dark red curve is a wavefront. (c) The unfolded setup of a longer distance case.}
\end{figure}
Here, we use the AWP to explain the amplitude and phase correlation pattern after propagation predicted by Chan \emph{et al.}\ \cite{PhysRevA.75.050101}. In most cases, $\sigma_+\gg\sigma_-$ (strong position correlation) or $\sigma_+\ll\sigma_-$ (strong anticorrelation). We choose the first case and approximate Eq.\ \eqref{dGwf} by
\begin{equation}
    \psi(\boldsymbol{\rho}_1,\boldsymbol{\rho}_2)\approx\exp\left(-\frac{|\boldsymbol{\rho}_1-\boldsymbol{\rho}_2|^2}{4\sigma_-^2}-\frac{|\boldsymbol{\rho}_2|^2}{\sigma_+^2}\right),
\end{equation}
and the equivalent classical system is a convolution by a narrow Gaussian function followed by a large Gaussian mask. In the unfolded setup, a point source propagates $z$, passes through such a system, and propagates $z$ again. Free-space propagation is also a convolution, so the light can go through the convolution part of the system before the first propagation, becoming a Gaussian beam with a small waist radius, as shown in Fig.\ \ref{dgproplfig}. If $z$ is smaller, the center of the resulting spatial state is approximately at the same transverse position; if $z$ is larger, it anticorrelates with the starting position. In either case, the phase pattern is roughly a spherical wave with the distance $2z$ centered by the starting position.

\section{Applications in practical protocols}

In this section, we apply the AWP to illustrate the QIUP and the quantum holography with polarization entanglement. For GI, an analysis of the spatial resolution and field of view influenced by the NLC thickness, which is often far less significant than the lens size and the pump beam width, is shown in Appendix \ref{AWPAQGI}.

\subsection{Quantum imaging with undetected photons}\label{QIUPsubsect}

\begin{figure}[t]
\includegraphics[width=0.46\textwidth]{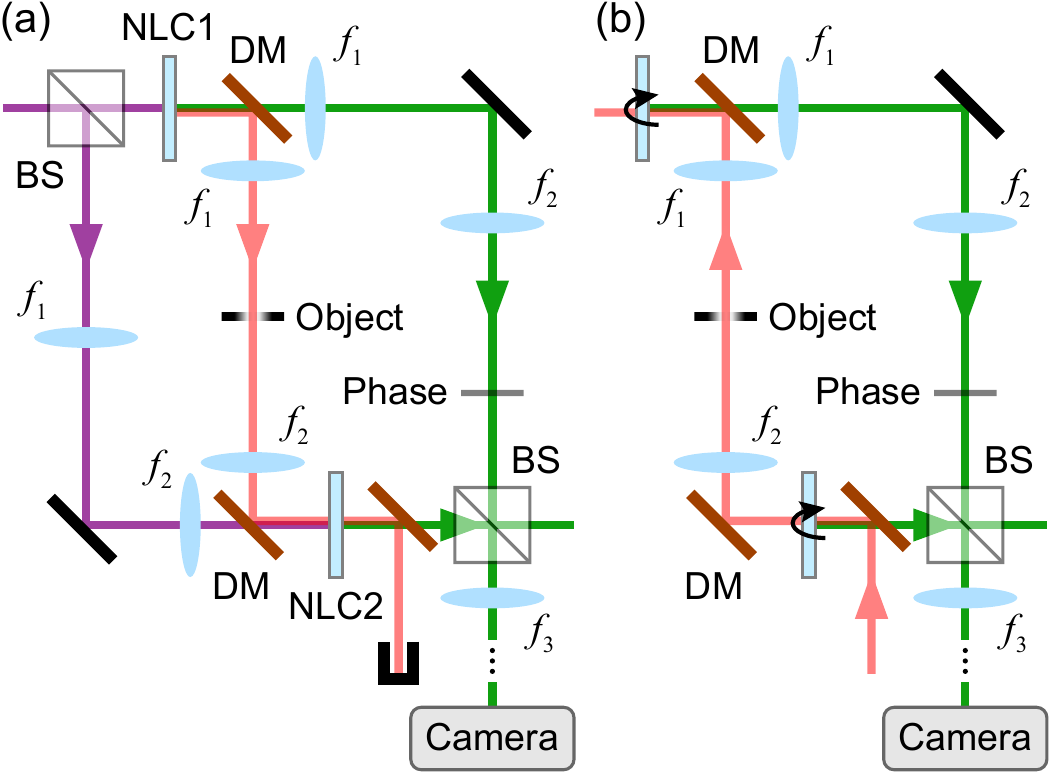}
\caption{\label{qiupfig}(a) The setup of Mach-Zehnder-type quantum imaging with undetected photons using momentum correlation. The pump beam is split into two beams which pump two identical nonlinear crystals (NLC1 and NLC2; after which the beams are not drawn). Using dichromic mirrors (DMs), the signal beam (green) from NLC1 with a constant phase interferes with the beam from NLC2 at the beam splitter (BS) and a camera detects one of the output ports. The idler beam (light red) from NLC1 passes through the object and is overlapped with the beam from NLC2 before being discarded. In the setup using position correlation, $f_1$, $f_2$, and $f_3$ are replaced by three $4f$ systems whose focal lengths are $f_1,f_2,\ldots,f_6$. (b) The AWP-unfolded setup of (a), where the NLCs reflect lights coming from the right side using a different wavelength, masks them according to the pump beam amplitude, and blurs them according to the crystal thickness. The absorptive part of the object also emits light. Different advanced wave sources have no coherence.}
\end{figure}

QIUP was proposed and experimentally demonstrated by Lemos \emph{et al.}\ in 2014 \cite{Lemos2014}, based on the induced coherence without induced emission phenomenon observed by Zou, Wang, and Mandel (ZWM) \cite{PhysRevLett.67.318,PhysRevA.44.4614}. There are two types of the setup, Mach-Zehnder (MZ) and Michelson \cite{BarretoLemos:22}, whose AWP-unfolded setups are indeed similar as MZ and Michelson interferometers and have no essential difference: the ``undetected'' \emph{i} light of unit intensity traveling backward passes through the second NLC, the object, and the first NLC (one NLC is used twice in the Michelson-type setup). When the object is absent, the detected \emph{s} lights reflected by the two crystals interfere at the beam splitter (BS), or only one \emph{s} beam from the second crystal shows no interference, so they can measure the complex $T(\boldsymbol{\rho})$ of thin objects. For simplicity, the two NLCs are identical with the same pump beam amplitude. QIUP is also categorized by the correlation types, position and momentum. The setup using momentum correlation and its AWP expansion are shown in Fig.\ \ref{qiupfig}. If the position correlation is used, the single Fourier lenses in the setup are replaced by $4f$ systems.

From the AWP theory with no detection (Appendix \ref{AWPND}), the object also emits \emph{i} light of the intensity $1-|T|^2$ (incoherent with the light emitted from the undetected path) to the first crystal \cite{otherawfromobject}, and the reflected \emph{s} light adds a background at the camera. If this contribution is omitted, one would obtain an incorrect interference visibility. For simplicity, we consider a single non-diffracting spatial mode (the original ZWM experiment with no lenses) at each path and let $T$ be real. Without the advanced wave from the object, the two fields with the amplitudes $1$ and $Te^{iC}$ interfere at the BS, the intensity at the camera is $(1+T^2+2T\cos C)/2$, and the visibility is $2T/(1+T^2)$ (which is for an MZ interferometer and the result if a detector at the undetected path triggers the camera). With the advanced wave of the intensity $1-T^2$ from the object which adds the background $(1-T^2)/2$, the visibility becomes the correct $T$ \cite{PhysRevLett.67.318,Lemos2014,BarretoLemos:22}.

Then, we consider the influence of the spatial resolution and the field of view by the NLCs with the Gaussian pump $U_p(\boldsymbol{\rho})=e^{-|\boldsymbol{\rho}|^2/w^2}$ in MZ-type QIUP with perfect lenses under the paraxial approximation. We do not analyze its interference visibility again, so the global coefficients are omitted in the following expressions of light field amplitudes.

\subsubsection{QIUP using momentum correlation}

In the unfolded setup of QIUP using momentum correlation, a point source emits the \emph{i} light at $\boldsymbol{\rho}_i$ of the second crystal $U_{i2}(\boldsymbol{\rho})=\delta(\boldsymbol{\rho}-\boldsymbol{\rho}_i)$, which immediately creates the reflected \emph{s} light whose amplitude is
\begin{equation}\label{us2prime}
    U_{s2'}(\boldsymbol{\rho})\approx\operatorname{Ssi}\left(\frac{2\pi|\boldsymbol{\rho}-\boldsymbol{\rho}_i|^2}{L\lambda_+}\right)U_p(\boldsymbol{\rho}_i).
\end{equation}
After the Fourier lens $f_3$, the amplitude at the camera plane is
\begin{align}\label{us3prime}
    U_{s3'}(\boldsymbol{\rho})&=\int d\boldsymbol{\rho}'U_{s2'}(\boldsymbol{\rho}')\exp\left(-\frac{i2\pi\boldsymbol{\rho}}{\lambda_sf_3}\cdot\boldsymbol{\rho}'\right)\nonumber\\
    &\propto\operatorname{sinc}\left(\frac{\pi L\lambda_+|\boldsymbol{\rho}|^2}{2\lambda_s^2f_3^2}\right)e^{-\frac{i2\pi\boldsymbol{\rho}_i}{\lambda_sf_3}\cdot\boldsymbol{\rho}}.
\end{align}
The \emph{i} light after the second crystal, the lens $f_2$, and the object has an amplitude $U_{i1}(\boldsymbol{\rho})=T(\boldsymbol{\rho})e^{-i2\pi\boldsymbol{\rho}_i\cdot\boldsymbol{\rho}/(\lambda_if_2)}$. After the lens $f_1$, it is
\begin{equation}
    U_{i0}(\boldsymbol{\rho})=\int d\boldsymbol{\rho}'T(\boldsymbol{\rho}')\exp\left[-\frac{i2\pi\boldsymbol{\rho}'}{\lambda_if_1}\cdot\left(\boldsymbol{\rho}+\frac{f_1\boldsymbol{\rho}_i}{f_2}\right)\right].
\end{equation}
From the alternative approximation in Sec.\ \ref{AWPESSPDCC}, this amplitude are multiplied by $U_p(\boldsymbol{\rho})$ and then convoluted by the Ssi term to approximately become the \emph{s} light amplitude
\begin{equation}
    U_{s0}(\boldsymbol{\rho})\approx[U_{i0}(\boldsymbol{\rho})U_p(\boldsymbol{\rho})]\ast\operatorname{Ssi}\left(\frac{2\pi|\boldsymbol{\rho}|^2}{L\lambda_+}\right),
\end{equation}
where ``$\ast$'' is the convolution sign. After the lens $f_1$, using the convolution theorem, the amplitude becomes
\begin{align}\label{us1} 
    &U_{s1}(\boldsymbol{\rho})=\int d\boldsymbol{\rho}'U_{s0}(\boldsymbol{\rho}')\exp\left(-\frac{i2\pi\boldsymbol{\rho}}{\lambda_sf_1}\cdot\boldsymbol{\rho}'\right)\nonumber\\
    \propto&\left\{\!\left[T\!\left(-\frac{\lambda_i\boldsymbol{\rho}}{\lambda_s}\right)e^{\frac{i2\pi\boldsymbol{\rho}_i}{\lambda_sf_2}\cdot\boldsymbol{\rho}}\right]\!\ast e^{-\frac{\pi^2w^2|\boldsymbol{\rho}|^2}{\lambda_s^2f_1^2}}\!\right\}\operatorname{sinc}\!\left(\frac{\pi L\lambda_+|\boldsymbol{\rho}|^2}{2\lambda_s^2f_1^2}\right)\!.
\end{align}
Now, the spatial resolution can be determined, which should be described by the width of an equivalent PSF acted at the \emph{object plane}, rather than the magnified or shrunk image. From Eq.\ \eqref{us1}, The PSF width is the FWHM of the squared Gaussian function $\sqrt{2\ln2}\lambda_sf_1/(\pi w)$, and the object is $-\lambda_s/\lambda_i$ times magnified, so the PSF width at the object plane (the spatial resolution) is $\sqrt{2\ln2}\lambda_if_1/(\pi w)$, independent of $\lambda_s$. Then, there is only a $4f$ system and a constant phase plate $e^{iC}$ (the phase from the reflection of the BS is absorbed into $C$), and
\begin{align}\label{us3}
    &U_{s3}(\boldsymbol{\rho})=e^{iC}U_{s1}(-f_2\boldsymbol{\rho}/f_3)\nonumber\\
    \approx\ &T\left(\frac{\lambda_if_2\boldsymbol{\rho}}{\lambda_sf_3}\right)\operatorname{sinc}\left(\frac{\pi L\lambda_+f_2^2|\boldsymbol{\rho}|^2}{2\lambda_s^2f_1^2f_3^2}\right)e^{-\frac{i2\pi\boldsymbol{\rho}_i}{\lambda_sf_3}\cdot\boldsymbol{\rho}+iC},
\end{align}
where the convolution is omitted as the spatial resolution has been analyzed. The magnification ratio is $\lambda_sf_3/(\lambda_if_2)$. Compared with Eq.\ \eqref{us3prime}, the tilt phase terms are the same, but the coefficients in the sinc terms are different. The narrower width determines the interference region and, thus, the field of view. When $f_2>f_1$, Eq.\ \eqref{us3} dominates, otherwise Eq.\ \eqref{us3prime} dominates, so the field of view is $1.882\lambda_i\operatorname{min}\{f_1,f_2\}/\sqrt{L\lambda_+}$. When the optical fields $[U_{s3}(\boldsymbol{\rho})+U_{s3'}(\boldsymbol{\rho})]/\sqrt{2}$ from point sources with different $\boldsymbol{\rho}_i$ values superpose incoherently, the relative intensity distribution does not change.

Finally, we discuss the spot size at the camera from the advanced wave emitted by the object. Letting the point source be at $\boldsymbol{\rho}_o$ of the object plane, we can borrow Eq.\ \eqref{us1}, replacing $T(\boldsymbol{\rho})$ by $\sqrt{1-|T(\boldsymbol{\rho}_o)|^2}\delta(\boldsymbol{\rho}-\boldsymbol{\rho}_o)$, and the blurring term has the same width, so the extra background does not affect the spatial resolution, which is also the same for QIUP using position correlation.

\subsubsection{QIUP using position correlation}

In QIUP using position correlation, we denote the absolute values of the magnification ratios of the three $4f$ systems as $M_2=f_2/f_1$, $M_4=f_4/f_3$, and $M_6=f_6/f_5$. With the point source at $\boldsymbol{\rho}_i$, the amplitude of the created \emph{s} light from the second crystal is still Eq.\ \eqref{us2prime}, which we denote by $U_{s4'}(\boldsymbol{\rho})$ in this case. So, at the camera plane,
\begin{align}\label{us6primeposition}
    U_{s6'}(\boldsymbol{\rho})&=U_{s4'}\left(-\frac{\boldsymbol{\rho}}{M_6}\right)=\operatorname{Ssi}\left(\frac{2\pi|\boldsymbol{\rho}+M_6\boldsymbol{\rho}_i|^2}{L\lambda_+M_6^2}\right)U_p(\boldsymbol{\rho}_i)\nonumber\\
    &\approx\operatorname{Ssi}\left(\frac{2\pi|\boldsymbol{\rho}+M_6\boldsymbol{\rho}_i|^2}{L\lambda_+M_6^2}\right)U_p\left(\frac{\boldsymbol{\rho}}{M_6}\right),
\end{align}
where the last line is from the fact that the Ssi term is narrow and $U_p(\boldsymbol{\rho})$ is an even function. At the object plane, the advanced wave amplitude $U_{i2}(\boldsymbol{\rho})=T(-\boldsymbol{\rho}_i/M_4)\delta(\boldsymbol{\rho}+\boldsymbol{\rho}_i/M_4)$. At the first crystal, $U_{i0}(\boldsymbol{\rho})=T(-\boldsymbol{\rho}_i/M_4)\delta[\boldsymbol{\rho}-\boldsymbol{\rho}_i/(M_2M_4)]$. So, the created \emph{s} light amplitude at this crystal is 
\begin{equation}\label{us0position} 
    U_{s0}(\boldsymbol{\rho})\approx T\!\left(\!-\frac{\boldsymbol{\rho}_i}{M_4}\!\right)\operatorname{Ssi}\left(\frac{2\pi\big|\boldsymbol{\rho}-\frac{\boldsymbol{\rho}_i}{M_2M_4}\big|^2}{L\lambda_+}\right)U_p\!\left(\frac{\boldsymbol{\rho}_i}{M_2M_4}\right)\!.
\end{equation}
After the three $4f$ systems and the constant phase plate,
\begin{align}\label{us6position} 
    &U_{s6}(\boldsymbol{\rho})=e^{iC}U_{s0}\left(-\frac{\boldsymbol{\rho}}{M_2M_4M_6}\right)\nonumber\\
    =\ &e^{iC}T\left(-\frac{\boldsymbol{\rho}_i}{M_4}\right)\operatorname{Ssi}\left(\frac{2\pi\big|\boldsymbol{\rho}+M_6\boldsymbol{\rho}_i\big|^2}{L\lambda_+M_2^2M_4^2M_6^2}\right)U_p\left(\frac{\boldsymbol{\rho}_i}{M_2M_4}\right)\nonumber\\
    \approx\ &e^{iC}T\!\left(\!\frac{\boldsymbol{\rho}}{M_4M_6}\!\right)\operatorname{Ssi}\!\left(\frac{2\pi\big|\boldsymbol{\rho}+M_6\boldsymbol{\rho}_i\big|^2}{L\lambda_+M_2^2M_4^2M_6^2}\right)U_p\!\left(\frac{\boldsymbol{\rho}}{M_2M_4M_6}\right)\!.
\end{align}
The magnification ratio is $M_4M_6$. From Eqs.\ \eqref{us6primeposition} and \eqref{us6position}, the coefficients of the Ssi terms are generally different, so the interference is not perfect unless $M_2M_4=1$. Then, the width of the pump beam term at the camera plane is $\sqrt{2\ln2}wM_6$, so the field of view is $\sqrt{2\ln2}wM_2$. The spatial resolution is $0.544M_2\sqrt{L\lambda_+}$. The field at the camera is a single spot, and an incoherent superposition of different $\boldsymbol{\rho}_i$ values forms the image.

Our results using the AWP, especially the dependence on the wavelengths and the pump beam width, are the same as the standard approach \cite{BarretoLemos:22,qute.202300353,Defienne2024}, but only classical-like optical fields are analyzed, which are much easier to understand. The dependence on the focal lengths and some coefficients may be different due to the choice of lens configurations and FWHM formulas.

\subsection{Quantum holography with polarization entanglement}\label{QHwPEsubsect}

\begin{figure}[t]
\includegraphics[width=0.38\textwidth]{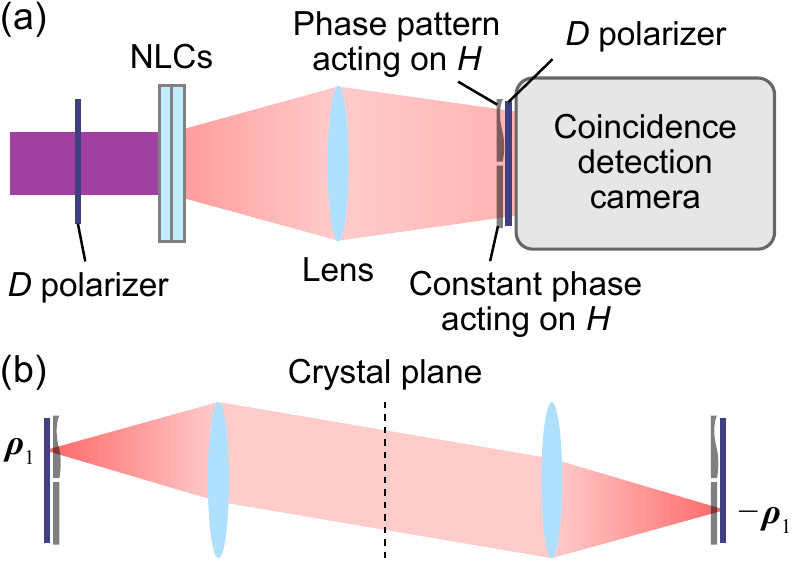}
\caption{\label{qhpefig}(a) The setup of the simplified quantum holography with polarization entanglement experiment. Two stacked identical NLCs produce biphotons entangled both in position and polarization when using a $D$-polarized pump beam. After a Fourier lens, the photons are position-anticorrelated. A phase pattern is added to the $H$ component of one half of the beam, and a constant phase is added to the other half. Then, the photons are $D$ polarized and coincidentally detected. (b) The AWP-unfolded setup under the thin-crystal approximation.}
\end{figure}

Quantum holography with polarization entanglement was realized by Defienne \emph{et al.}\ in 2021 \cite{Defienne2021}, while Camphausen \emph{et al.}\ performed a similar experiment \cite{sciadv.abj2155}. A simplified setup is shown in Fig.\ \ref{qhpefig}(a), whose standard interpretation is that a diagonally ($D$) polarized beam pumps two stacked thin NLCs rotated by $90^\circ$ with respect to each other (for example, the first crystal has a constant $\chi_{VHH}^{(2)}$ value, and $\chi_{HVV}^{(2)}$ of the second crystal has the same value), producing degenerate collinear type-I down-converted biphotons with position and polarization entanglement. After a Fourier lens, the position-anticorrelated biphotons can be regarded as distinguishable by dividing the beam into two halves. Using a spatial light modulator (SLM) which only modulates the phases of $H$ lights, photon 1 and 2 are modulated by a phase pattern $e^{i\Phi(\boldsymbol{\rho})}$ and the constant phases $C=0,\pi/2,\pi,3\pi/2$, respectively, and the state is
\begin{equation}
    \int d\boldsymbol{\rho}[e^{i\Phi(\boldsymbol{\rho})+iC}|H\rangle_1|\boldsymbol{\rho}\rangle_1|H\rangle_2|-\boldsymbol{\rho}\rangle_2+|V\rangle_1|\boldsymbol{\rho}\rangle_1|V\rangle_2|-\boldsymbol{\rho}\rangle_2]
\end{equation}
in the ideal case. Then, they are both projected to $D$ polarization and the JPD of two anticorrelated positions $\boldsymbol{\rho},-\boldsymbol{\rho}$ is
\begin{equation}\label{peqhjpd}
    |e^{i\Phi(\boldsymbol{\rho})+iC}+1|^2\propto1+\cos[\Phi(\boldsymbol{\rho})+C],
\end{equation}
ready for phase-shifting holography \cite{Yamaguchi:97}.

In the AWP theory including polarization, if the NLCs are thin, pumped by a $D$-polarized normally incident plane wave, the $H$ and $V$ components of light stay the same after the crystal plane in the unfolded setup, so nothing is placed at this plane. Otherwise, a polarizer should exist if there is only one crystal. In another word, as the spatially entangled state $\int d\boldsymbol{\rho}|\boldsymbol{\rho}\rangle|\boldsymbol{\rho}\rangle$ reflects the spatial state perfectly, the similar polarization state $(|H\rangle|H\rangle+|V\rangle|V\rangle)/\sqrt{2}$ ($H$ and $V$ can be replaced by any two orthogonal linear polarizations) reflects the polarization. So, when designing new quantum imaging protocols, to preserve the polarization in the unfolded setup, the SPDC source should be such a stacked NLC pair pumped by a $D$ beam. If the pump beam has another linear polarization, the creation amplitudes of the $H$ and $V$ lights are different. If it is partially polarized or unpolarized, similar as the partially coherent pump beam case discussed in Appendix \ref{AWPPCPB}, the result is an incoherent superposition of two states using pump lights with orthogonal polarizations.

In the ideal case, the unfolded setup is a $D$ light source (at one half of the original beam at the camera), an SLM with $\Phi(\boldsymbol{\rho})$, a $4f$ system, an SLM with $C$, and a $D$ polarizer, as shown in Fig.\ \ref{qhpefig}(b). Letting the focal length of the Fourier lens be $f$, a $D$ polarized point source at $\boldsymbol{\rho}_1$ described by $\delta(\boldsymbol{\rho}-\boldsymbol{\rho}_1)|D\rangle$ becomes $\delta(\boldsymbol{\rho}+\boldsymbol{\rho}_1)[e^{i\Phi(\boldsymbol{\rho}_1)+iC}|H\rangle+|V\rangle]$ after the $4f$ system and the two SLMs, which is at the position $-\boldsymbol{\rho}_1$. Projected to $D$ polarization, the resulting intensity is Eq.\ \eqref{peqhjpd}. In real experiments, the spatial resolution is not limited by the physical principle as the object is close to the camera, and the crystal thickness affects the field of view. The center planes of the two NLCs are different, which means the equivalent diffraction distances (see Sec.\ \ref{IRFoOS}) between the two lenses in the unfolded setup are different for the $H$ and $V$ light. Letting the thickness of a single NLC be $L$, the difference $2L/n_o$ leads to a paraboloid phase difference between the $H$ and $V$ components at the camera plane even when $\Phi(\boldsymbol{\rho})=0$, which should be compensated for by the SLM.

\section{Discussion}\label{discusssec}

In this article, beyond the thin-crystal case, based on low-gain SPDC, we developed a formal and general theory of the AWP with monochromatic pump, signal, and idler lights, reducing the biphoton optical problem to a classical one except the possible polarization or wavelength change which does not affect the linearity. A bulk SPDC crystal can approximated by a spatial filtering system in the AWP-unfolded setup. Then, some biphoton phenomena including the free-space propagation of the double-Gaussian state and some quantum imaging protocols, although they are intrinsically quantum, can be interpreted in an equivalent classical setup, which will be helpful to the design of novel biphoton devices \cite{pnas.1714936115,PhysRevX.11.031044} including interference, holography \cite{Belinskii1994,Abouraddy:01holo}, and imaging, as well as analyzing properties and limits of existing techniques. In our analyses, the paraxial approximation was still used for convenience \cite{PhysRevResearch.4.033252}, but more realistic diffraction theories, including the Rayleigh-Sommerfeld or the vector theory \cite{nanooptics}, can be chosen. For the numerical calculation of the biphoton wave function, the expression in Eq.\ \eqref{awpcore} is the common formula of the standard approach and the AWP, so the calculation of the whole wave function cannot be simplified. However, if we are only interested in the conditional wave functions of a few points, they can be handled using existing calculation methods of classical light propagation, and properties of the biphoton state can be quickly determined for realistic applications.

An interesting question is, using SPDC and some linear, passive optical systems, whether an arbitrary biphoton transverse state $\psi(\boldsymbol{\rho}_1,\boldsymbol{\rho}_2)$ can be prepared. The wave function corresponds to the impulse response function of a classical optical setup $h(\boldsymbol{\rho},\boldsymbol{\rho}_0)=\psi(\boldsymbol{\rho}_0,\boldsymbol{\rho})$ in the AWP, so, in the nondegenerate SPDC case, after designing a classical setup with a plane separating two regions adapted for the two wavelengths, a thin NLC can be placed at this plane, and the folded quantum optical setup produces the desired biphoton state. In the degenerate case, the two photons pass through the same setup after a thin NLC, and the AWP-unfolded setup must be $\hat{A}\hat{\Lambda}\hat{\bar{A}}=\hat{A}\hat{\Lambda}\hat{A}^T$ using the matrix notation (Appendix \ref{OSaQO}), where the diagonal $\hat{\Lambda}$ corresponds to the pump beam amplitude. As $h(\boldsymbol{\rho},\boldsymbol{\rho}_0)$ is also symmetric, its matrix can be decomposed into $\hat{U}\hat{\Lambda}\hat{U}^T$, where $\hat{U}$ is unitary and $\hat{\Lambda}$ is real, nonnegative, and diagonal, exactly corresponding to this case. So, the question is reduced to whether an arbitrary linear optical process (lossless or not) in classical optics can be realized \cite{schmidtprep,LopezPastor:21,Kulce2021}. Beyond the case of a thin NLC, the utilization of other nonlinear materials \cite{PhysRevLett.117.123901,Rozenberg:22,adma.202313589,meta2023} for a more effective biphoton state preparation can be studied in the future.

The HOM effect \cite{PhysRevLett.59.2044,pnas.2010827117,Bouchard_2021} is that two identical photons arriving at the two input ports of a balanced BS respectively at the same time must occupy the same output path. It can measure the optical path difference (OPD) if the photons have finite temporal widths, and has been applied in OPD imaging \cite{Ndagano2022}. Using the monochromatic AWP, if we consider only one non-diffracting spatial mode at each path, it can only explain why two photons with the same polarization cannot exit at different paths: the unfolded setup is a Sagnac interferometer (the NLC becomes a mirror), where a beam entering one port of the BS must come out from the same port, no matter whether there is an OPD between the two paths. If the photon pairs are nondegenerate, as the wavelength is changed in the Sagnac ring, an OPD causes the two beams propagating clockwise and counterclockwise to have a phase difference, leading to the beating phenomenon \cite{PhysRevLett.61.54}. In Ref.\ \cite{KLYSHKO1988133}, Klyshko briefly discussed the HOM effect after suggesting the AWP which includes the finite wave packet lengths. The main idea is the detected position emits a pulse traveling \emph{backward in time} at the time of detection, which often passes through a bandpass interference filter widening the pulse, and arrives at the NLC. The reflected wave packet (the retarded wave) travels forward in time as the conditional biphoton state. Appendix \ref{PTQMCAWP} is a theory of quasimonochromatic AWP with an illustration of the HOM effect. A formal theory of the AWP for arbitrary polychromatic lights using a quantization method for light fields in arbitrary media \cite{PhysRevA.53.1818,PhysRevA.57.3931,PhysRevLett.117.123901} is beneficial in interpreting and inventing novel quantum optical techniques with pulsed SPDC photons.

\section*{Acknowledgments}

This work was supported by the Innovation Program for Quantum Science and Technology (No.\ 2021ZD0301200 and No.\ 2021ZD0301400), National Natural Science Foundation of China (No.\ 11821404 and No.\ 92365205), and USTC Major Frontier Research Program (No.\ LS2030000002).
    
\appendix

\section{Green's function of diffraction}\label{appGFD}

In the angular spectrum theory, LHI medium propagation means a paraboloid phase $k_z(\mathbf{q})z$ is added to the angular spectrum, where
\begin{equation}\label{kzo}
k_z(\mathbf{q})=\sqrt{|\mathbf{k}|^2-|\mathbf{q}|^2}\approx nk-|\mathbf{q}|^2/(2nk)
\end{equation}
under the paraxial approximation, and $\mathbf{k}$ is the wave vector in the medium. The inverse Fourier transform of $e^{ik_z(\mathbf{q})z}$ is the Green's function in Eq.\ \eqref{greeniso}.

For a uniaxial birefringent crystal with the refractive indices $n_o$ and $n_e$ whose optical axis lies in the $x$-$z$ plane \cite{WALBORN201087,Karan_2020}, letting the angle between the optical axis and the $z$ axis be $\theta$, we have the walk-off coefficient
\begin{equation}
    \alpha=\frac{(n_o^2-n_e^2)\tan\theta}{n_o^2\tan^2\theta+n_e^2},
\end{equation}
the effective refractive index of a normally incident plane wave
\begin{equation}
    \eta=\frac{n_on_e}{\sqrt{n_o^2\sin^2\theta+n_e^2\cos^2\theta}},
\end{equation}
$\beta=\eta^2/(n_on_e)$, $\gamma=\eta/n_e$, and, for an \emph{e} light,
\begin{equation}\label{kze}
    k_z(\mathbf{q})\approx\eta k-\frac{\beta^2k_x^2+\gamma^2k_y^2}{2\eta k}+\alpha k_x.
\end{equation}
The inverse Fourier transform yields
\begin{equation}\label{greenbirefbetagamma}
    G_{e,L}(\boldsymbol{\rho})=\frac{\eta e^{i\eta kL}}{i\beta\gamma\lambda L}\exp\left\{i\frac{\eta k}{2L}\left[\frac{(x+\alpha L)^2}{\beta^2}+\frac{y^2}{\gamma^2}\right]\right\}.
\end{equation}
Note that $\beta\approx\gamma\approx1$, so it can be further approximated to Eq.\ \eqref{greenbiref}. which is similar as an isotropic medium with the refractive index $\eta$ except for the displacement $-\alpha L\mathbf{e}_x$.

\section{Optical systems as quantum operators}\label{OSaQO}

When we are only interested in the transverse optical field at a given $z$, we use $|\boldsymbol{\rho}\rangle$ to describe the state with the created photon at a single position $\hat{a}_{k,\boldsymbol{\rho},z}^\dagger|\mathrm{vac}\rangle$. Denoting the photon state at the input plane as $|\psi\rangle=\int d\boldsymbol{\rho}U(\boldsymbol{\rho})|\boldsymbol{\rho}\rangle$ and the optical system as a quantum operator (or a matrix in the position basis)
\begin{equation}
    \hat{A}=\int d\boldsymbol{\rho}d\boldsymbol{\rho}_0h(\boldsymbol{\rho},\boldsymbol{\rho}_0)|\boldsymbol{\rho}\rangle\langle\boldsymbol{\rho}_0|,
\end{equation}
we have $h(\boldsymbol{\rho},\boldsymbol{\rho}_0)=\langle\boldsymbol{\rho}|\hat{A}|\boldsymbol{\rho}_0\rangle$, and the transverse state at the output plane is simply $\hat{A}|\psi\rangle$. $\hat{A}$ is unitary only when the system is lossless, including any paraxial free-space propagation and infinitely large pure phase objects. Denoting the inverse optical system of $\hat{A}$ as $\hat{\bar{A}}$, the reciprocity theorem is $\hat{\bar{A}}=\hat{A}^T$, which is preserved when two reciprocal systems are combined $\hat{\bar{A}}_1\hat{\bar{A}}_2=\hat{A}_1^T\hat{A}_2^T=(\hat{A}_2\hat{A}_1)^T$.

For thin objects $h(\boldsymbol{\rho},\boldsymbol{\rho}_0)=T(\boldsymbol{\rho}_0)\delta(\boldsymbol{\rho}-\boldsymbol{\rho}_0)$ and propagation processes in a bulk LHI media $h(\boldsymbol{\rho},\boldsymbol{\rho}_0)=G_L(\boldsymbol{\rho}-\boldsymbol{\rho}_0)$ [$G_L(\boldsymbol{\rho})$ is an even function], the functions are symmetric and so are the corresponding operators $\hat{A}=\hat{A}^T$. Without knowing the reciprocity theorem, if we assert the operators apply to both the lights traveling forward and backward $\hat{\bar{A}}=\hat{A}$, these two systems are reciprocal, and so is a combination of them.

Now we use operators to derive the better-known AWP. Under the thin-crystal approximation, the created biphoton state is $\int d\boldsymbol{\rho}U_p(\boldsymbol{\rho})|\boldsymbol{\rho}\rangle|\boldsymbol{\rho}\rangle$. When photons 1 and 2 pass through two systems $\hat{A}_1$ and $\hat{A}_2$, respectively, the transverse wave function
\begin{align}
    &\langle\boldsymbol{\rho}_1|\langle\boldsymbol{\rho}_2|\hat{A}_1\hat{A}_2\int d\boldsymbol{\rho}U_p(\boldsymbol{\rho})|\boldsymbol{\rho}\rangle|\boldsymbol{\rho}\rangle\nonumber\\
    =&\int d\boldsymbol{\rho}\langle\boldsymbol{\rho}_2|\hat{A}_2U_p(\boldsymbol{\rho})|\boldsymbol{\rho}\rangle\langle\boldsymbol{\rho}|\hat{\bar{A}}_1|\boldsymbol{\rho}_1\rangle=\langle\boldsymbol{\rho}_2|\hat{A}_2\hat{T}\hat{\bar{A}}_1|\boldsymbol{\rho}_1\rangle,
\end{align}
where $\hat{T}=\int d\boldsymbol{\rho}U_p(\boldsymbol{\rho})|\boldsymbol{\rho}\rangle\langle\boldsymbol{\rho}|$ can be regarded as a thin object masking the field. So, $\hat{A}_2\hat{T}\hat{\bar{A}}_1$ is the system in the unfolded setup.

\section{Standard SPDC analysis in the momentum space}\label{SPDCAMS}

For degenerate type-I SPDC of a bulk crystal, the biphoton momentum wave function at $z=0$ is \cite{WALBORN201087,Karan_2020}
\begin{equation}\label{generalspdcmwf} 
    \tilde{\psi}(\mathbf{q}_1,\mathbf{q}_2)=\tilde{U}_p(\mathbf{q}_p)\operatorname{sinc}\!\left\{\frac{L}{2}\big[k_{p,z}(\mathbf{q}_p)\!-\!k_z(\mathbf{q}_1)\!-\!k_z(\mathbf{q}_2)\big]\!\right\}\!,
\end{equation}
where $\tilde{U}_p(\mathbf{q}_p)$ is the angular spectrum of the pump beam at $z=0$ and $\mathbf{q}_p=\mathbf{q}_1+\mathbf{q}_2$. Substituting Eq.\ \eqref{kzo} for the down-converted light with the vacuum wave number $k$ and Eq.\ \eqref{kze} for the pump light ($\beta\approx\gamma\approx1$) into Eq.\ \eqref{generalspdcmwf}, a simplified form cannot be obtained generally, so we consider the following special cases.

If the pump beam is a normally incident plane wave $\tilde{U}(\mathbf{q})=\delta(\mathbf{q})$, then $\mathbf{q}_2$ must be $-\mathbf{q}_1$, and the sinc term in Eq.\ \eqref{generalspdcmwf} becomes
\begin{equation}
    \operatorname{sinc}\left[(\eta-n_o)kL+\frac{L|\mathbf{q}_1-\mathbf{q}_2|^2}{8n_ok}\right],
\end{equation}
which becomes Eq.\ \eqref{standardmspdccol} when $\eta$ of the pump light equals $n_o$ of the down-converted light. Otherwise, noncollinear phase matching satisfies when $\eta<n_o$ and no phase matching can be satisfied when $\eta>n_o$, and the position wave function cannot be easily written.

If the pump beam is not a plane wave, the collinear phase matching is satisfied $\eta=n_o$, and the walk-off of the \emph{e} pump beam is ignored $\alpha\approx0$, the sinc term in Eq.\ \eqref{generalspdcmwf} becomes $\operatorname{sinc}[L|\mathbf{q}_1-\mathbf{q}_2|^2/(8n_ok)]$, and the approximated momentum wave function is \cite{PhysRevLett.92.127903,PhysRevA.95.063836,Schneeloch_2016}
\begin{equation}\label{colapproxspdcmwf}
    \tilde{\psi}(\mathbf{q}_1,\mathbf{q}_2)=\tilde{U}_p(\mathbf{q}_1+\mathbf{q}_2)\operatorname{sinc}\left(\frac{L|\mathbf{q}_1-\mathbf{q}_2|^2}{8n_ok}\right).
\end{equation}
Considering the formula ($a>0$)
\begin{equation}\label{ssisincx2}
    \frac{1}{(2\pi)^2}\int d\mathbf{q}\operatorname{sinc}(a|\mathbf{q}|^2)e^{i\mathbf{q}\cdot\boldsymbol{\rho}}=-\frac{1}{4\pi a}\operatorname{Ssi}\left(\frac{|\boldsymbol{\rho}|^2}{4a}\right),
\end{equation}
its inverse Fourier transform is \cite{PhysRevA.95.063836}
\begin{equation}\label{colapproxspdcwf}
    \psi(\boldsymbol{\rho}_1,\boldsymbol{\rho}_2)=U_p\left(\frac{\boldsymbol{\rho}_1+\boldsymbol{\rho}_2}{2}\right)\operatorname{Ssi}\left(\frac{n_ok|\boldsymbol{\rho}_1-\boldsymbol{\rho}_2|^2}{2L}\right).
\end{equation}

Finally, for the nondegenerate collinear type-I case pumped by a normally incident plane wave, $k_p=k_s+k_i$, $\eta k_p=n_sk_s+n_ik_i$, and $\mathbf{q}_i$ still equals $-\mathbf{q}_s$. The momentum wave function is
\begin{equation}\label{nondegenmwf}
    \tilde{\psi}(\mathbf{q}_s,\mathbf{q}_i)=\delta(\mathbf{q}_s+\mathbf{q}_i)\operatorname{sinc}\left(\frac{L\lambda_+|\mathbf{q}_s-\mathbf{q}_i|^2}{32\pi}\right).
\end{equation}
So, the momentum width or the position correlation strength is dependent on the wavelengths of the \emph{s} and \emph{i} lights inside the crystal. If the pump beam has a finite size, the expression becomes complicated. We should also let the pump beam be wide enough (its angular spectrum is much narrower than the sinc term) so that $\mathbf{q}_i\approx-\mathbf{q}_s$ and $\alpha\approx0$, and obtain
\begin{equation}\label{colapproxspdcmwfnd}
    \tilde{\psi}(\mathbf{q}_s,\mathbf{q}_i)\approx\tilde{U}_p(\mathbf{q}_s+\mathbf{q}_i)\operatorname{sinc}\left(\frac{L\lambda_+|\mathbf{q}_s-\mathbf{q}_i|^2}{32\pi}\right),
\end{equation}
which is used in Sec.\ \ref{AWPESSPDCC}.

\section{AWP of other cases}\label{AWPothercases}

\subsection{Case of partially coherent pump beam}\label{AWPPCPB}

If the classical pump beam is partially coherent in transverse space, it must be quasimonochromatic. Similar as the density matrix in quantum mechanics, the mutual coherence function (at the same time) can be decomposed into an incoherent superposition of various coherent light fields \cite{PhysRevA.62.043816}
\begin{equation}
    J_p(\mathbf{r},\mathbf{r}')=\sum_jU_{pj}(\mathbf{r})U_{pj}^\ast(\mathbf{r}'),
\end{equation}
where the probabilities are absorbed into the amplitudes. Letting $\hat{D}_j$ be a product of many single-mode displacement operators which produces the pump light $U_{pj}(\mathbf{r})$, i.e., $\hat{a}_\mathbf{r}\hat{D}_j|\mathrm{vac}\rangle=U_{pj}(\mathbf{r})\hat{D}_j|\mathrm{vac}\rangle$, in the standard approach, the initial state is a mixture of pure states with the pump mode $U_{pj}(\mathbf{r})$ and the vacuum signal and idler modes $\sum_j\hat{D}_j|\mathrm{vac}\rangle\langle\mathrm{vac}|\hat{D}_j^\dagger$. Then, the two-photon component of the down-converted state
\begin{align}
    \hat{\rho}=\sum_j&\int d\mathbf{r}\chi^{(2)}(\mathbf{r})U_{pj}(\mathbf{r})\hat{a}_{k_1,\mathbf{r}}^\dagger\hat{a}_{k_2,\mathbf{r}}^\dagger|\mathrm{vac}\rangle\nonumber\\
    &\times\langle\mathrm{vac}|\int d\mathbf{r}'\chi^{(2)}(\mathbf{r}')U_{pj}^\ast(\mathbf{r}')\hat{a}_{k_1,\mathbf{r}'}\hat{a}_{k_2,\mathbf{r}'},
\end{align}
and the second-order correlation function
\begin{align}\label{SOCFmix}
    &G^{(2)}(\mathbf{r}_1,\mathbf{r}_2,\mathbf{r}'_1,\mathbf{r}'_2)=\operatorname{Tr}\Big(\hat{a}_{\mathbf{r}_1}\hat{a}_{\mathbf{r}_2}\hat{\rho}\hat{a}_{\mathbf{r}'_1}^\dagger\hat{a}_{\mathbf{r}'_2}^\dagger\Big)\nonumber\\
    =\ &\sum_j\int d\mathbf{r}\chi^{(2)}(\mathbf{r})U_{pj}(\mathbf{r})h_1(\mathbf{r}_1,\mathbf{r})h_2(\mathbf{r}_2,\mathbf{r})\nonumber\\
    &\quad\ \ \times\int d\mathbf{r}'\chi^{(2)}(\mathbf{r}')U_{pj}^\ast(\mathbf{r}')h_1^\ast(\mathbf{r}'_1,\mathbf{r}')h_2^\ast(\mathbf{r}'_2,\mathbf{r}')\nonumber\\
    =\ &\sum_j\psi_j(\mathbf{r}_1,\mathbf{r}_2)\psi_j^\ast(\mathbf{r}'_1,\mathbf{r}'_2),
\end{align}
where $\psi_j(\mathbf{r}_1,\mathbf{r}_2)$ is the biphoton wave function from the pump beam component $U_j(\mathbf{r})$. So, in the AWP, the state of photon 2 when photon 1 is postselected to $\mathbf{r}_1$ [letting $\mathbf{r}'_1=\mathbf{r}_1$ in Eq.\ \eqref{SOCFmix}] is an incoherent superposition of the photon 2 states from the $\mathbf{r}_1$ source when the pump beam takes all the components. If one is only interested in the JPD, the JPDs from all the components $|\psi_j(\mathbf{r}_1,\mathbf{r}_2)|^2$ are directly summed.

For example, biphotons created by a spatially incoherent pump beam using a sufficiently thin NLC have no momentum correlation. The reason is that the unfolded setup of a Fourier lens after the crystal is a point source at $\boldsymbol{\rho}_1$ and a $4f$ system centered by the crystal plane. The decomposition of such a pump beam is point sources all over the beam spot. Using each component, a small hole is placed at the crystal plane, so the intensity distribution at the output plane is wide, independent of the hole position or $\boldsymbol{\rho}_1$. The summed intensity over different hole positions as the conditional probability distribution is also independent of $\boldsymbol{\rho}_1$.

\subsection{Case of bucket detection}\label{AWPBD}

In the case of bucket detection of photon 1 at a surface region $S_1$, the photon 2 states with different photon 1 positions superpose incoherently, and the resulting state is generally mixed. With the initial state $|\Psi\rangle$ [Eq.\ \eqref{spdcstate}], the conditional state after photon 1 bucket detection is $\hat{\rho}_\mathrm{buc}=\int_{S_1}d\mathbf{r}_1\hat{a}_{\mathbf{r}_1}|\Psi\rangle\langle\Psi|\hat{a}_{\mathbf{r}_1}^\dagger$, and the mutual coherence function (corresponding to the first-order correlation function) of photon 2 is
\begin{align}\label{bucketmcf}
    J_2(\mathbf{r}_2,\mathbf{r}'_2)&=\operatorname{Tr}\Big(\hat{a}_{\mathbf{r}_2}\hat{\rho}_\mathrm{buc}\hat{a}_{\mathbf{r}'_2}^\dagger\Big)\nonumber\\
    &=\int_{S_1}d\mathbf{r}_1\langle\mathrm{vac}|\hat{a}_{\mathbf{r}_1}\hat{a}_{\mathbf{r}_2}|\Psi\rangle\langle\Psi|\hat{a}_{\mathbf{r}_1}^\dagger\hat{a}_{\mathbf{r}'_2}^\dagger|\mathrm{vac}\rangle\nonumber\\
    &=\int_{S_1}d\mathbf{r}_1\psi(\mathbf{r}_1,\mathbf{r}_2)\psi^\ast(\mathbf{r}_1,\mathbf{r}'_2),
\end{align}
which is an incoherent superposition of the photon 2 state when photon 1 is postselected to different points at $S_1$. If we are only interested in the intensity of photon 2, the result is $\int_{S_1}d\mathbf{r}_1|\psi(\mathbf{r}_1,\mathbf{r}_2)|^2$. In the AWP, treating the biphoton wave function as an impulse response function, the expression has the form of van Cittert-Zernike theorem, and the bucket detector becomes an extended light source with constant intensity. If the polarization is considered, the source emits advanced waves of two orthogonal polarizations incoherently like a natural light.

\subsection{Case of no detection}\label{AWPND}

If photon 1 is undetected, photon 2 may also be present even when photon 1 is absorbed by the medium \cite{PhysRevLett.117.123901}. Phenomenologically, $\hat{a}_{k,\mathbf{r}}^\dagger$ for photon 1 can have an additional term describing the internal excitation $\hat{b}_{\mathbf{r}}^\dagger$ of the absorptive media
\begin{equation}
    \hat{a}_{k,\mathbf{r}}^\dagger=\int d\mathbf{r}'h(\mathbf{r}',\mathbf{r})\big[\hat{a}_{\mathbf{r}'}^\dagger+\gamma(\mathbf{r}')\hat{b}_{\mathbf{r}'}^\dagger\big],
\end{equation}
where $\gamma(\mathbf{r})$ is a real function related to the absorption at the photon 1 wavelength at $\mathbf{r}$. For photon 2, the old $\hat{a}_{k,\mathbf{r}}^\dagger$ can still be used, as the $\hat{b}_{\mathbf{r}}^\dagger$ term is undetectable. However, the two operators $\hat{a}_{\mathbf{r}}^\dagger$ and $\hat{b}_{\mathbf{r}}^\dagger$ are extremely unequal. With a given source point, the impulse response function generally extends to infinity and cannot be normalized. If a closed surface $S$ covering the whole setup blocks all the light at the photon 1 wavelength, an infinite amount of field outside $S$ disappears, becoming the internal excitation of the finite absorber, and $\gamma(\mathbf{r})$ should be very large in its nonzero regions. One can also calculate the mutual coherence function of photon 2
\begin{align}\label{undetectedallmcf}
    J_2(\mathbf{r}_2,\mathbf{r}'_2)&=\langle\Psi|\hat{a}_{\mathbf{r}'_2}^\dagger\hat{a}_{\mathbf{r}_2}|\Psi\rangle\nonumber\\
    &=\int d\mathbf{r}_1\psi(\mathbf{r}_1,\mathbf{r}_2)\psi^\ast(\mathbf{r}_1,\mathbf{r}'_2)[1+\gamma^2(\mathbf{r}_1)],
\end{align}
which means points all over the space emit advanced waves and superpose incoherently, and the significant contributions are only from the infinite space outside $S$ and the absorptive regions inside $S$.

Then, we assume photon 1 created inside $S$ is either absorbed by the media inside $S$ or detected by a bucket detector covering $S$ (without which it travels to infinity). Thus, we do not consider setups with perfect cavities. Then, we have
\begin{align}\label{undetectedSmcf}
    J_2(\mathbf{r}_2,\mathbf{r}'_2)=&\int_S d\mathbf{r}_1\psi(\mathbf{r}_1,\mathbf{r}_2)\psi^\ast(\mathbf{r}_1,\mathbf{r}'_2)\nonumber\\
    &+\int d\mathbf{r}_1\gamma^2(\mathbf{r}_1)\psi(\mathbf{r}_1,\mathbf{r}_2)\psi^\ast(\mathbf{r}_1,\mathbf{r}'_2),
\end{align}
which means the state of photon 2 is from an incoherent superposition of advanced waves from points at $S$ of unit intensity, as well as those from the absorptive regions inside $S$ with the intensity $\gamma^2(\mathbf{r})$. Now, the advanced-wave sources do not extend to infinity, and the impact of absorption on the optical field of interest is no longer infinite.

Considering an antireflective coated absorptive medium with the refractive index $n$ and the absorptive coefficient $\alpha$ (in this subsection, $\alpha$ is not the walk-off coefficient) at $0<z<L$, the amplitude of a plane wave traveling along $\mathbf{e}_z$ is $Ce^{-\alpha z/2+in\omega z/c}/\sqrt{n}$ (the field amplitude before the input plane $C$ is real) inside the medium, and the amplitude transmittance $T=e^{-\alpha L/2}$ (ignoring its phase). Imagine that a small region $d\mathbf{r}$ reflects $\alpha dz$ of the energy passing through it to an imaginary space, where the reflected light is detected in the vacuum. In the AWP, a source in the imaginary space emits the advanced wave of unit intensity, which is reflected to the real space at $\mathbf{r}$ with the intensity $\alpha dz$. Advanced waves from different $\mathbf{r}$ points superpose incoherently, and the total intensity at the left surface $z=0$ is $\int_0^Ldz\alpha e^{-\alpha z}=1-e^{-\alpha L}=1-T^2$. So, if photon 1 is undetected, the advanced wave intensity from a thin semitransparent object is $1-T^2$, and the object is similar as a BS with the amplitude transmittance $T$ whose reflected path is detected by a bucket detector.

Finally, we need to find the expression of $\gamma(\mathbf{r})$. In the case above, it should be a constant $\gamma$. The classical intensity is proportional to $n|E|^2$ inside the medium, while the wave functions are based on the electric field in the main text, so $\gamma=\sqrt{\alpha/n}$ to ensure the source intensity is $\alpha$.

\subsection{Case of $N$ photons}\label{AWPNP}

The AWP can describe quantum optical fields with more than two photons. In spontaneous parametric $N$-photon generation involving the $N$th-order nonlinearity,
\begin{equation}\label{spdcNstate}
    |\Psi\rangle=\int d\mathbf{r}\chi^{(N)}(\mathbf{r})U_p(\mathbf{r})\prod_{j=1}^N\hat{a}_{k_j,\mathbf{r}}^\dagger|\mathrm{vac}\rangle,
\end{equation}
and
\begin{equation} 
    \psi(\mathbf{r}_1,\ldots,\mathbf{r}_N)=\!\int\!d\mathbf{r}\!\prod_{j\neq j_0}^N\!h_j(\mathbf{r},\mathbf{r}_j)\chi^{(N)}(\mathbf{r})U_p(\mathbf{r})h_{j_0}(\mathbf{r}_{j_0},\mathbf{r}).
\end{equation}
So, the conditional wave function of one photon with the positions of all other photons given is from the product of the amplitudes of the other photons propagating backward, the $N$th-order susceptibility, and the pump beam amplitude.

For example, with a sufficiently thin NLC pumped by a normally incident plane wave, if photon $j$ ($j\neq1$) 
with the wave number $k_j$ propagates a distance $d_j$ and is postselected to the center point $\mathbf{0}$, in the AWP, its amplitude at the crystal is $e^{ik_j|\boldsymbol{\rho}|^2/(2d_j)}$, and the amplitude of photon 1 at the crystal is a spherical wave with the distance $k_1/\sum_{j=2}^N(k_j/d_j)$ \cite{PhysRevA.76.045802}.

\section{Integral in degenerate collinear type-I SPDC}\label{CalcInt}

From the second line of Eq.\ \eqref{distwfdci}, treating $z$ as a constant, the integral over $\boldsymbol{\rho}$ is
\begin{align}
    &\int d\boldsymbol{\rho}\exp\left[{-\frac{in_ok}{2z}(|\boldsymbol{\rho}-\boldsymbol{\rho}_1|^2+|\boldsymbol{\rho}_2-\boldsymbol{\rho}|^2)}\right]\nonumber\\
    \propto&\ z\exp\left(-\frac{in_ok}{4z}|\boldsymbol{\rho}_1-\boldsymbol{\rho}_2|^2\right).
\end{align}
Denoting $a=n_ok|\boldsymbol{\rho}_1-\boldsymbol{\rho}_2|^2/4$, then we calculate the Cauchy principal value of the integral over $z$:
\begin{equation}
    \int_{-L/2}^{L/2}dz\frac{\exp(-ia/z)}{z}=\int_{-L/2}^{L/2}dz\frac{\cos(a/z)-i\sin(a/z)}{z}.
\end{equation}
$\cos(a/z)/z$ is an odd function, so its integral is zero. Letting $\zeta=1/z$, then
\begin{align}
    &\int_{-L/2}^{L/2}dz\frac{\sin(a/z)}{z}=\left(\int_{-L/2}^{-0}+\int_{+0}^{L/2}\right)dz\frac{\sin(a/z)}{z}\nonumber\\
    =&\left(\int_{-\infty}^{-2/L}+\int_{2/L}^{+\infty}\right)d\zeta\frac{\sin(a\zeta)}{\zeta}=-2\operatorname{Ssi}(2a/L).
\end{align}
So, the integration result is proportional to Eq.\ \eqref{standardpspdccol}.

\section{Calculation of degenerate beamlike type-II SPDC}\label{CDBLT2SPDC}

``Beamlike'' type-II ($e\to o+e$) SPDC was named and realized by Takeuchi \cite{Takeuchi:01} in 2001, while Kurtsiefer \emph{et al.}\ also performed a similar experiment \cite{PhysRevA.64.023802}. In this configuration, the \emph{o} and \emph{e} down-converted lights are in two separate beams in the far field, suitable for biphoton collection. We let photon 1 be the \emph{o} light, and the biphoton wave function is the first line of Eq.\ \eqref{distwfdci} with the pump beam amplitude reverted to $e^{i2\eta kz}$ (the value of $\eta$ is determined later by eliminating the $e^{iaz}$ term) and $G_{-z}(\boldsymbol{\rho}_2-\boldsymbol{\rho})$ replaced by $G_{e,-z}(\boldsymbol{\rho}_2-\boldsymbol{\rho})$,
\begin{align}\label{distwfdbii}
    &\psi(\boldsymbol{\rho}_1,\boldsymbol{\rho}_2)=\int_\textrm{crystal}d\mathbf{r}G_{-z}(\boldsymbol{\rho}-\boldsymbol{\rho}_1)e^{i2\eta kz}G_{e,-z}(\boldsymbol{\rho}_2-\boldsymbol{\rho})\nonumber\\
    \propto&\int d\boldsymbol{\rho}\int_{-L/2}^{L/2}\frac{dz}{z^2}e^{i(\eta-n_o)kz-\frac{ik}{2z}\left(n_o|\boldsymbol{\rho}-\boldsymbol{\rho}_1|^2+\eta|\boldsymbol{\rho}_2-\boldsymbol{\rho}-\alpha z\mathbf{e}_x|^2\right)}\nonumber\\
    \propto&\int_{-L/2}^{L/2}\frac{dz}{z}\exp\left[-\frac{in_o\eta\alpha k(x_1-x_2)}{n_o+\eta}-\frac{in_o\eta k|\boldsymbol{\rho}_1-\boldsymbol{\rho}_2|^2}{2(n_o+\eta)z}\right]\nonumber\\
    \propto&\exp\left[i\theta_\textrm{beam}k(x_2-x_1)\right]\operatorname{Ssi}\left[\frac{n_o\eta k|\boldsymbol{\rho}_1-\boldsymbol{\rho}_2|^2}{L(n_o+\eta)}\right],
\end{align}
where $\eta=n_o(\alpha^2+\sqrt{16+\alpha^4})/4>n_o$ and $\theta_\textrm{beam}=n_o\eta\alpha/(n_o+\eta)$. Compared with the collinear type-I case, the wave function has a different width and a tilt phase term, so the beams are displaced in the momentum space.

If such a state is double-Gaussian approximated and propagates a distance in the free space, compared to the result in Sec.\ \ref{PdGs}, the initial Gaussian beam in the unfolded setup is tilted by $\theta_\textrm{beam}$.

\section{AWP analysis of quantum ghost imaging}\label{AWPAQGI}

GI was first experimentally demonstrated in 1995 \cite{ghostimaging} together with ghost interference \cite{PhysRevLett.74.3600}. The well-known unfolded setups are simply an imaging system for GI and a Young's interference experiment for ghost interference. The bucket detector can be placed immediately after the object (while a focusing lens is usually used in experiments). In the AWP, if the point source is at the opaque (semitransparent) region of the object, it is immediately totally (partially) absorbed and the corresponding position at the image plane has no (a weaker) spot. Summing the resulting optical fields at the image plane incoherently over all the source points yields the object image. Here, we discuss the influence of the spatial resolution of GI from the NLC thickness when the imaging lens is perfect with a magnification ratio $-M$ ($M>0$).

\begin{figure}[t]
\includegraphics[width=0.48\textwidth]{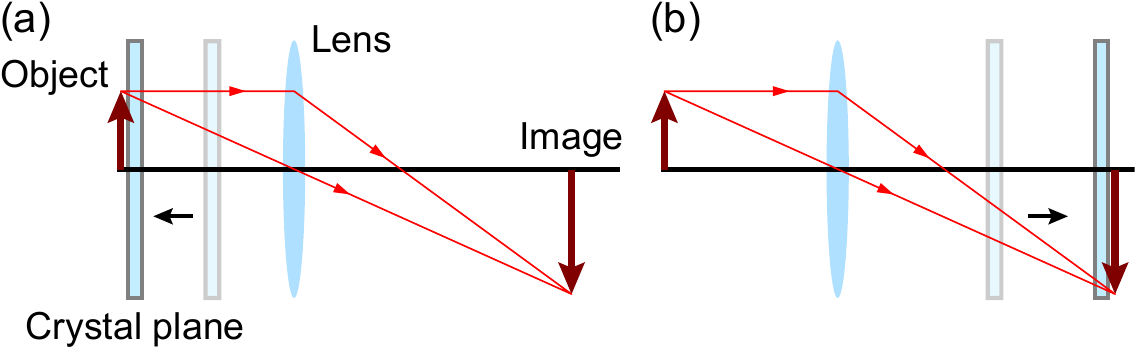}
\caption{\label{gifig}In the unfolded setup of ghost imaging, the crystal plane (which should be an amplitude masking and spatial filtering system, but it is drawn as an actual crystal) can be moved in the free space. (a) When the lens is at the detection path, to analyze the spatial resolution caused by the crystal thickness, the crystal plane is moved close to the object plane. (a) When the lens is at the object path, the crystal plane is moved close to the image plane.}
\end{figure}
In the degenerate case, collinear type-II SPDC was originally used \cite{ghostimaging}, where the biphotons have the form of two rings in the momentum space and are surely correlated in position at the crystal, but the wave function possibly cannot be written in a simple form. Collinear type-I SPDC can be used, but a beam splitter (BS) should be inserted, which means half the photons are lost. We consider this case here, and the pump beam is a normally incident plane wave. If the lens is placed at the spatially resolved detection path, the crystal plane is between the object and the lens in the unfolded setup, as shown in Fig.\ \ref{gifig}(a). As free-space propagation and the convolution from the crystal can be exchanged, the crystal plane can be moved to the object plane, and the convolution is performed before propagation. The PSF width is clearly $\Delta\rho_\mathrm{d}$. If the lens is at the object path, the crystal plane can be moved to the image plane in the unfolded setup, as shown in Fig.\ \ref{gifig}(b), and the PSF width is $\Delta\rho_\mathrm{d}/M$.

In the nondegenerate collinear type-I case, let the \emph{s} light pass through the object in the original setup. If the crystal plane is between the object and the lens in the unfolded setup, considering the form of spherical waves with different wavelengths, compared to the degenerate $\lambda_i$ case, the distance between the object and the crystal is multiplied by $\lambda_i/\lambda_s$ (to preserve the transverse amplitude at the crystal plane) and the magnification ratio is unchanged. The propagation processes from the object to the crystal and from the crystal to the lens are still two convolutions, so the convolution from the crystal can be performed first, and the PSF width is $\Delta\rho_\mathrm{nd}$. If the crystal is at the detection path, compared to the degenerate $\lambda_s$ case, the distance between the crystal and the image plane is multiplied by $\lambda_s/\lambda_i$, and the convolution from the crystal can be the last step, so the PSF width is $\Delta\rho_\mathrm{nd}/M$.

If the pump beam width or the lens size is finite, irises are inserted to the unfolded setup, whose impact on the spatial resolution \cite{PhysRevA.72.013810,PhysRevA.78.033836} and the field of view can also be analyzed in a classical manner. In many GI setups, rather than the biphoton correlation strength, these two factors are actually the major cause of image blurring.

\section{Quasimonochromatic AWP}\label{PTQMCAWP}

\subsection{Theory}

We consider quasimonochromatic \emph{s} and \emph{i} lights, which pass through two hypothetical bandpass filters with the bandwidth $\Delta\omega$ and the central angular frequencies $\bar{\omega}_s$ and $\bar{\omega}_i$, respectively ($\Delta\omega\ll\bar{\omega}_s,\bar{\omega}_i$), immediately after their creation. The transmission functions of the filters are rectangular, so passing through it multiple times has no difference from passing once \cite{perfectrect}. Ignoring the $\sqrt{\omega(\mathbf{k})}$ term in the electric field operator, we use $\hat{a}_{s,\mathbf{r}}^\dagger$ ($\hat{a}_{i,\mathbf{r}}^\dagger$) to describe the creation of the \emph{s} (\emph{i}) light at $\mathbf{r}$ which is immediately filtered, and $\hat{a}_{s,\mathbf{r}}$ ($\hat{a}_{i,\mathbf{r}}$) to describe the ordinary annihilation operator which only applies on the \emph{s} (\emph{i}) light (there is only $\hat{a}_\mathbf{r}$ in the degenerate case). For example,
\begin{equation}
    \hat{a}_{s,\mathbf{r}}^\dagger=\int_{\bar{\omega}_s-\frac{\Delta\omega}{2}}^{\bar{\omega}_s+\frac{\Delta\omega}{2}}d\omega\hat{a}_{\omega,\mathbf{r}}^\dagger=\int_{\bar{\omega}_s-\frac{\Delta\omega}{2}}^{\bar{\omega}_s+\frac{\Delta\omega}{2}}d\omega\int d\mathbf{r}'h_\omega(\mathbf{r}',\mathbf{r})\hat{a}_{\mathbf{r}'}^\dagger,
\end{equation}
where $\hat{a}^\dagger_{\omega,\mathbf{r}}$ means $\hat{a}^\dagger_{\omega/c,\mathbf{r}}$ in the main text. Then, we define the quasimonochromatic impulse response function of a stationary linear optical system. If a photon with the \emph{s} light spectrum is created at $\mathbf{r}_0$, let $h_s(\mathbf{r},\mathbf{r}_0,t)$ be the spatial wave function at $\mathbf{r}$ after the time $t$ ($t\geq0$), which should satisfy
\begin{equation}
    h_s(\mathbf{r},\mathbf{r}_0,t)\propto\int_{\bar{\omega}_s-\frac{\Delta\omega}{2}}^{\bar{\omega}_s+\frac{\Delta\omega}{2}}d\omega h_\omega(\mathbf{r},\mathbf{r}_0)e^{-i\omega t},
\end{equation}
as the down-converted lights are linear after creation. By applying the reciprocity to $h_\omega(\mathbf{r},\mathbf{r}_0)$, we have $h_s(\mathbf{r}_0,\mathbf{r},t)=h_s(\mathbf{r},\mathbf{r}_0,t)$. There is a similar definition for the \emph{i} light.

Letting $\hat{\mathcal{U}}(t)=e^{-i\hat{H}_0t/\hbar}$ be the time evolution operator applied to both the \emph{s} and \emph{i} light, where the explicit form of $\hat{H}_0$ which depends on the optical setup is unnecessary and may be non-Hermitian [if the system is absorptive; then $\hat{\mathcal{U}}(t)$ is not unitary], we have $\hat{\mathcal{U}}(t)|\mathrm{vac}\rangle=|\mathrm{vac}\rangle$ and
\begin{align}
    \langle\mathrm{vac}|\hat{a}_{s,\mathbf{r}_s}\hat{a}_{i,\mathbf{r}_i}\hat{\mathcal{U}}(t)\hat{a}_{s,\mathbf{r}_{s0}}^\dagger\hat{a}_{i,\mathbf{r}_{i0}}^\dagger|\mathrm{vac}\rangle\nonumber\\
    =h_s(\mathbf{r}_s,\mathbf{r}_{s0},t)h_i(\mathbf{r}_i,\mathbf{r}_{i0},t).
\end{align}
We assume $\chi^{(2)}(\mathbf{r})$ has no dispersion in our considered spectrum. With the pump light $U_p(\mathbf{r},t)$ and
\begin{equation}
    \hat{H}'(t)=\int d\mathbf{r}C\chi^{(2)}(\mathbf{r})U_p(\mathbf{r},t)\hat{a}_{s,\mathbf{r}}^\dagger\hat{a}_{i,\mathbf{r}}^\dagger+\mathrm{H.c.},
\end{equation}
where H.c.\ means the Hermitian conjugate of the last expression and $C$ is a constant, the effective Hamiltonian is time-dependent $\hat{H}(t)=\hat{H}_0+\hat{H}'(t)$. Letting $t_0$ be an initial time (which can be $-\infty$) so that $\chi^{(2)}(\mathbf{r})U_p(\mathbf{r},t)=0$ when $t\leq t_0$, $t$ be the observation time, and $\Delta t$ be an infinitesimal time interval, under the first-order approximation, we have
\begin{align}
    |\Psi(t)\rangle&=\left\{\prod_{t'=t_0}^{t}\left[1-\frac{i}{\hbar}\hat{H}_0\Delta t-\frac{i}{\hbar}\hat{H}'(t')\Delta t\right]\right\}|\mathrm{vac}\rangle\nonumber\\
    &\approx\left\{\prod_{t'=t_0}^{t}\left[\hat{\mathcal{U}}(\Delta t)-\frac{i\Delta t}{\hbar}\hat{\mathcal{U}}(\Delta t)\hat{H}'(t')\right]\right\}|\mathrm{vac}\rangle\nonumber\\
    &\approx\left[1-\frac{i\Delta t}{\hbar}\sum_{t'=t_0}^{t}\hat{\mathcal{U}}(t-t')\hat{H}'(t')\right]|\mathrm{vac}\rangle\nonumber\\
    &=|\mathrm{vac}\rangle-\frac{i}{\hbar}\int_{t_0}^tdt'\hat{\mathcal{U}}(t-t')\hat{H}'(t')|\mathrm{vac}\rangle,
\end{align}
where the product and sum signs add $\Delta t$ to $t'$ each time. The biphoton wave function at $t$ is \cite{Belinskii1994}
\begin{align} 
    &\psi(\mathbf{r}_s,\mathbf{r}_i,t)=\langle\mathrm{vac}|\hat{a}_{s,\mathbf{r}_s}\hat{a}_{i,\mathbf{r}_i}|\Psi(t)\rangle\nonumber\\
    =&-\frac{iC}{\hbar}\int_{t_0}^tdt'\int d\mathbf{r}\chi^{(2)}(\mathbf{r})U_p(\mathbf{r},t')\nonumber\\
    &\qquad\qquad\qquad\quad\ \times\langle\mathrm{vac}|\hat{a}_{s,\mathbf{r}_s}\hat{a}_{i,\mathbf{r}_i}\hat{\mathcal{U}}(t-t')\hat{a}_{s,\mathbf{r}}^\dagger\hat{a}_{i,\mathbf{r}}^\dagger|\mathrm{vac}\rangle\nonumber\\
    \propto&\int_{t_0}^t\!dt'\!\int\!d\mathbf{r}\chi^{(2)}(\mathbf{r})U_p(\mathbf{r},t')h_s(\mathbf{r}_s,\mathbf{r},t-t')h_i(\mathbf{r}_i,\mathbf{r},t-t'),
\end{align}
which is from the contributions of biphoton states created at all the moments from $t_0$ to $t$. Using the reciprocity, it is also written as
\begin{equation}
    \int_{t_0}^tdt'\int d\mathbf{r}\chi^{(2)}(\mathbf{r})U_p(\mathbf{r},t')h_s(\mathbf{r},\mathbf{r}_s,t-t')h_i(\mathbf{r}_i,\mathbf{r},t-t').
\end{equation}
\begin{figure}[t]
\includegraphics[width=0.45\textwidth]{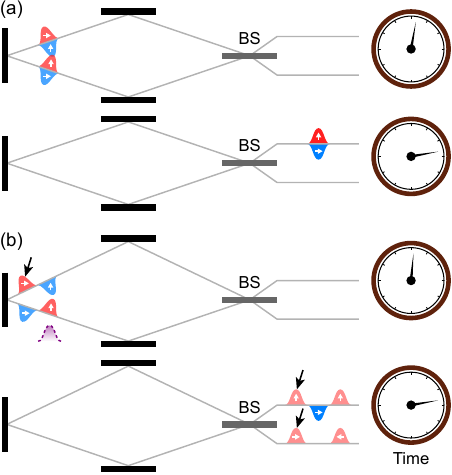}
\caption{\label{homfig}The wave packets traveling backward (blue, advanced, below the gray lines) and forward (red, retarded, above the gray lines) in time at different moments in the AWP of an HOM interferometer, if an advanced wave is emitted at the upper output path. The arrow in each wave packet indicate its phase as a unit vector at the complex plane. Their opacities indicate the amplitudes. (a) The case of a balanced HOM interferometer. (b) An unbalanced HOM interferometer with a longer upper path. The purple wave packet and black arrows are used to illustrate the case of pulsed pump light (see the text).}
\end{figure}
So, in the quasimonochromatic AWP, when evaluating the biphoton wave function at $t$, letting a point source at $\mathbf{r}_s$ emit the \emph{s} light at this time, the light field ``after'' $t-t'$ is $h_s(\mathbf{r},\mathbf{r}_s,t-t')$, but it travels backward in time (the original meaning of the advanced wave), arriving at $t'$, when the field amplitude is multiplied by $\chi^{(2)}(\mathbf{r})U_p(\mathbf{r},t')$ as the creation amplitude of the \emph{i} light at $t'$ which then travels forward in time (the retarded wave) by $t-t'$. Summing the contributions of all $t'$ values from $t_0$ to $t$ coherently yields the wave function. The probability of two photons being detected at $\mathbf{r}_s$ and $\mathbf{r}_i$ in coincidence at $t$ is proportional to $|\psi(\mathbf{r}_s,\mathbf{r}_i,t)|^2$.

\subsection{Application in Hong-Ou-Mandel interferometer}

In the HOM interferometer with a single non-diffracting spatial mode at each path, we follow the convention that the light reflected by the BS gains a $\pi/2$ phase. Letting the temporal wave packet be an even function $f(z)$, and the upper path be $d$ longer than the lower path. Using a monochromatic pump beam, when the upper output path detects a photon at $z_1$, the amplitude of the other photon at the same time is $i[f(z-z_1-d)+f(z-z_1+d)]/2$ at the upper path and $[-f(z-z_1-d)+f(z-z_1+d)]/2$ at the lower path in the AWP. As illustrated in Fig.\ \ref{homfig}, when $d=0$, the other photon must be at the same path; when $d$ is larger than the wave packet length, it may be at the other path and there may be a coincidence count, as the detectors cannot precisely measure the arrival time.

If the pump light is pulsed, as shown in the dashed purple wave packet in Fig.\ \ref{homfig}(b), waves pointed at by black arrows disappear. If the advanced wave is created at the same position but a different time, and neither of its components at the two paths overlaps with the pump wave when it backtracks to the crystal, there are no retarded waves. So, at a certain time, the two down-converted photons are at localized regions, while they can be anywhere (still correlated) before detection if the pump light is monochromatic.


\begin{thebibliography}{97}
\bibitem [{rho()}]{rhoorr}%
  \BibitemOpen
  \href@noop {} {\bibinfo {title}
  {{$\bar{h}_\mathrm{ps}(\mathbf{r},\mathbf{r}_0)$ is applicable only when
  $\mathbf{r}$ is at $S_1$ and $\mathbf{r}_0$ is at $S_\mathrm{ps}$. The
  two-dimensional $\boldsymbol{\rho}$ should have been used instead of
  $\mathbf{r}$, but these functions are defined at given planes which might not
  be parallel to the $x$-$y$ plane.}}}\BibitemShut {Stop}%
\bibitem [{8fs()}]{8fsystem}%
  \BibitemOpen
  \href@noop {} {\bibinfo {title} {{Letting the focal lengths of the four
  Fourier lenses be $f$ [$f>L/(2n_o)$], the first Fourier lens should be placed
  at $z=f+L(1-1/n_o)/2$ so that the optical field propagating forward at $z=0$
  is projected to the back focal plane $z=z_b=8f+L(1-1/n_o)/2$ with a constant
  phase $kL(n_o-1/n_o)/2$ which can be omitted.}}}\BibitemShut {Stop}%
\bibitem [{rep()}]{replace12}%
  \BibitemOpen
  \href@noop {} {\bibinfo {title} {{$h_-(\mathbf{r},\mathbf{r}_1)$ and
  $h_+(\mathbf{r}_2,\mathbf{r})$ should be replaced by
  $h_{1-}(\mathbf{r},\mathbf{r}_1)$ and $h_{2+}(\mathbf{r}_2,\mathbf{r})$,
  respectively, in nondegenerate SPDC.}}}\BibitemShut {Stop}%
\bibitem [{sch()}]{schmidtprep}%
  \BibitemOpen
  \href@noop {} {\bibinfo {title} {{Besides the AWP, the arbitrary biphoton
  state preparation problem can also be analyzed by Schmidt decomposition,
  whose coefficients correspond to the transmittance distribution of the
  amplitude mask, and the two unitary processes transform the position
  eigenstates to the two Schmidt bases (the same basis if the photons are
  indistinguishable).}}}\BibitemShut {Stop}%
\bibitem [{per()}]{perfectrect}%
  \BibitemOpen
  \href@noop {} {\bibinfo {title} {{If so, the position wave function of a
  photon has a width immediately at the time of creation, violating the
  causality. Thus, actual bandpass filters introduce delays and cannot have a
  perfect rectangular transmission function, whose Fourier transform, the sinc
  function, extends to infinity. This issue is ignored here.}}}\BibitemShut
  {Stop}%
\end{thebibliography}
\end{document}